\documentstyle[aaspp4,psfig]{article} 
\begin{document}

\title{Near-infrared Imaging and [O~I] spectroscopy of IC 443 using 2MASS and ISO}
\author{Jeonghee Rho, T. H. Jarrett, R. M. Cutri, and W. T. Reach
} 
\affil{Infrared Processing and Analysis Center, California Institute of
 Technology, MS 100-22, Pasadena, CA, 91125}

\centerline{Accepted for publication in The Astrophysical Journal} 
\centerline{Feburary 2001 issue}

\def\simlt{\lower.5ex\hbox{$\; \buildrel < \over \sim \;$}}

\centerline{For higher quality of figures, see http://spider.ipac.caltech.edu/staff/rho/}
\begin{abstract}

We present near-infrared J (1.25$\mu$m), H (1.65$\mu$m), and K$_s$
(2.17$\mu$m) imaging of the entire supernova remnant IC 443 from the Two Micron
All Sky Survey (2MASS), and Infrared Space Observatory (ISO) LWS
observations of [O~I] for 11 positions in the northeast. Near-infrared
emission from IC 443 was detected in all three bands from most of the
optically bright parts of the remnant, revealing a shell-like
morphology, with bright K$_s$ band  emission along the southern ridge
and bright J and H along the northeastern rim. 
The total
luminosity within the 2MASS bands is 1.3$\times$ 10$^{36}$ erg
s$^{-1}$.
These data represent the first near-infrared images that are complete
in coverage of the remnant.
 
The color and morphological structure are very different between the
northeastern and southern parts.  J and H band emission from the
northeast rim is comparably bright and  can be explained mostly by
[Fe~II] line emission.  The hydrogen recombination lines, P$\beta$ and
Br10, should also be present in the broad-band images, but probably
contribute less than 10\% of the J and H band fluxes.  Strong [O~I]
(63$\mu$m) lines  were detected crossing the northeastern rim, with the
strongest line in the northeastern shell where the near-infrared emission shows
filamentary structure. In contrast, the southern ridge is
dominated by K$_s$ band light exhibiting a clumped and  knotty
structure. 
A two excitation temperature model derived 
from previous ISO and ground-based observations
predicts that H$_2$ lines can explain most of K$_s$ band 
and at least half of J and H band emission.  
Hence, the prominent broad-band color differences arise from physically
different mechanisms: atomic fine structure lines along the
northeastern rim and molecular ro-vibrational lines along the southern
ridge. Shock models imply a fast J-shock with v$_s$ $\sim$ 100 km
s$^{-1}$ and 10$<$n$_o$ $<$10$^3$ cm$^{-3}$ for the northeastern
rim and  a slow
C-shock with  v$_s$ $\sim$ 30 km s$^{-1}$ and n$_o$ $\sim$ 10$^{4}$
cm$^{-3}$ for the southern ridge, respectively.

The shocked H$_2$ line emission  is well known from the southern sinuous
ridge, produced by an interaction with dense molecular clouds. The
large field of view and color of the 2MASS images show that the K$_s$
band emission extends to the east and the northeast, suggesting  that
the interaction extends to the inner part of the northeastern shell.
Our new CO map of the inner part of the northeast quadrant shows good
correspondence with the K$_s$-band map. The CO lines are broad,
confirming that the K$_s$-band emission is due to shocked H$_2$.

\noindent{\it Subject headings}: infrared:shock waves - 
nebula:individual (IC 443) - ISM:supernova remnants

\end{abstract}

\section{Introduction}

IC 443 is the best-studied case of supernova remnant-molecular cloud
interaction. 
The system is observed to have broad molecular lines with
CO gas accelerated to 20 km s$^{-1}$ and  shocked H$_2$ emission along
the southern ridge (van Dishoeck et al. 1993; Burton et al.
1988). The molecular cloud stretches across the remnant from the north
to the south as projected along the line of sight  (Cornett, Chin \&
Knapp 1977). IC 443 is in the Gem OB 1 association, located close to
the Galactic plane (G189.1+3.0).  
Optical spectroscopy of the northeastern filaments of IC
443 indicated a shock velocity of v$_s$$\sim$65-100 km s$^{-1}$ and a
preshock density of n$_o$$\sim$10-20 cm$^{-3}$ (Fesen \& Kirshner
1980). Shocked H~I is associated with the bright northeastern filaments
and extends across the entire southern hemisphere of the remnant with
velocities up to 100 km s$^{-1}$  (Giovanelli \& Haynes 1979).  IC 443
has unusual morphology; the optical and radio morphology is shell-like, but
it extends eastward beyond the main shell (referred to
the ``northeastern arm"). 
The optical emission shows
radial filaments with different curvature in the southwest,
and the X-ray
emission appears both at the interior of the shell and 
at the edge of the shell 
(Petre et al. 1988; Rho et al. 1993). 
A superposition with another SNR 
G189.6+3.3 has been suggested (Asaoka \& Aschenbach 1994).

The presence of H$_2$ line emission along the southern ridge strongly
suggests that the supernova shock wave is interacting with the molecular
cloud  (Burton et al. 1988; Richter et al. 1995).  A
near-infrared line of [Fe~II] (1.64$\mu$m) is detected from    the
northeastern shell, 30 times as strong as that of Br$\gamma$, which is
much stronger than seen in Galactic H~II regions. The strong
[Fe~II] line is produced not only by efficient excitation of Fe but
also by grain destruction (Graham et al. 1987).

Infrared emission from supernova remnants, including IC 443, is not
well understood.  IRAS observations show bright emission from IC 443,
which  was interpreted as thermal continuum 
emission from heated small and large dust grains  (Arendt 1989; Saken
et al. 1992; Braun \& Strom 1986). However, recent ISO observations
show no continuum between 5-14$\mu$m with a noise level of
0.2$\times$$10^{-4}$  erg s$^{-1}$ cm$^{-2}$ sr$^{-1}$
(Cesarsky et al. 1999).
Instead, the observed mid-infrared
spectrum of the shocked gas is dominated by rich emission lines
of H$_2$. 
ISO SWS observations of the northeastern filaments (Oliva et al. 1999a)
also showed detections of rich emission lines,
not molecular but ionic lines such as [Ne~II], 
[Fe~II] and [Si~II].
Previous interpretations of the IRAS 12 and 25$\mu$m
emission are different, based on differences between the ISO SWS  
and ISOCAM results.  
In particular, Oliva et al. (1999a) suggested that 25$\mu$m IRAS
emission is virtually all ionized ion lines, while Cesarsky et al
(1999) claimed that shocked H$_2$ lines  (at least at the southern region)
are dominant. This is due to limited field of view of ISO observations
toward IC 443, and also due to limited spatial resolution of the 
IRAS image.  

In this paper, we present J, H, and K$_s$ imaging from the Two Micron All
Sky Survey (2MASS, Skrutskie et al. 1997) covering the 
entire size of the IC 443 remnant, and we 
show which spectral lines  
contribute to the J, H, and K$_s$ band emission.
We present detections of the [O~I] 63$\mu$m line using ISO LWS observations. 
Combining the 2MASS and ISO data,  we employ  diagnostics for rating
the cooling in atomic or molecular shocks, and compare the line
brightnesses with  J and C shock models.

While previous near-infrared imaging and spectroscopic observations
were limited to a part of the supernova remnant, the 2MASS image
offers a global view of the emission, and it provides a
means to understand the surrounding ISM environment. 
IC 443 is the
first such example, and  we  expect  other supernova
remnants will be revealed in the near-infrared light using 2MASS.

\section{Observation}

\subsection{Two Micron All Sky Survey}
Using identical telescopes in the northern and southern hemispheres,
2MASS is mapping the entire sky in the J (1.11-1.36$\mu$m), H (1.5-1.8$\mu$m)
and K$_s$ (2-2.32$\mu$m) bands to a limiting point source sensitivity of
approximately 16.5, 16.0 and 15.5 magnitudes, respectively.  2MASS
observed IC 443 on 1997, November 23 UT using the northern telescope
at Mt. Hopkins, Arizona.  The data are acquired in
8.5$'$$\times$6$^{\circ}$ $``$tiles" for net 7.8 sec exposures of 
each point on the sky,
and pipeline data processing
(Cutri et al. 2000) produces calibrated 8.5$'$x17$'$ 
Atlas Images in 
the three survey bands. 
The image resolution is typically 3.5$''$, twice 
the camera resolution, and the data are sampled 
with 1$''$ per pixel.

A total of twenty-four 2MASS Atlas Images in each band covering the ~1$^{\circ}
\times 1^{\circ}$ region  around IC 443  were combined to produce the
large mosaic shown in Figure 1.  The mosaic was constructed by first
combining in the declination direction the three  Atlas Images (per
band) that fall in the same 2MASS tile  into individual strips.  Eight
such strips were formed in each band.   The low frequency atmospheric
emission gradients in each 2MASS $``$tile" were then fit out,
using a 17$'$ median 
filter in the declination direction.
This background subtraction doesn't affect the nebular emission 
scales with which this work is concerned, and any large scale 
($>$17$'$)
emission of the SNR is obscured by the low frequency atmospheric emission
gradients.
The background
levels in each pixel of the eight strips were then adjusted using  the
intensity offsets in the approximately 1$'$ RA overlap region between
tiles.
Surface brightness profiles in Figure 2 show the
resultant background fluctuation
in the right ascension direction,
and the average surface brightnesses over a 1 arcmin$^2$ area
for a representative position from the northeastern rim for J and H bands
and from the southern ridge for K$_s$ band  
are also shown.

The
2MASS flux calibrations for each band were 
7.46$\times 10^{-6}$Jy/DN,
5.75$\times 10^{-6}$Jy/DN and 7.12$\times 10^{-6}$Jy/DN, for J, H and
K$_s$, respectively.
These values are derived using the zero point
flux of 1603, 1075, and 698 Jy, and 
the average zero point magnitudes (background) are 
20.83, 20.68, and 20.0 magnitude, 
for J, H and K$_s$, respectively.
The current absolute zero point fluxes of 2MASS  
have an estimated 5\% uncertainty. 
The filter FWHM passbands are 0.25, 0.3, and 0.32$\mu$m,
J, H and K$_s$, respectively, where the long wavelength
cutoff in J band is defined by atmospheric H$_2$O and CO$_2$ 
absorption (Cutri et al. 2000). 
The central wavelength of 1.25, 1.65, and 2.17$\mu$m, for J, H and
K$_s$, respectively, is used.  
The unit DN is equivalent to  
1.52$\times 10^{-4}$, 0.83$\times 10^{-4}$, and
0.61$\times 10^{-4}$ erg s$^{-1}$ cm$^{-2}$ sr$^{-1}$, 
for J, H, and K$_s$ band, respectively, which are used to
estimate the surface brightness shown in Table 1.

\subsection{2MASS Near-Infrared Images}

Emission from IC 443 was detected in all 3 bands from most of the
optically bright parts of the remnant,  revealing a shell-like
morphology. The K$_s$ band image also 
shows bright emission at the southern ridge.  
These are the first near-infrared 
images that cover the entire remnant. 
Figure 1 shows the combined J, H and K$_s$ mosaic image, with
J assigned to blue, H to green and K$_s$ to red color.
The surface brightness in Figure 1 ranges from (0.3
- 6.4), (0.2 - 4.0), (0.1 - 12) with units of $10^{-4}$ erg s$^{-1}$
cm$^{-2}$sr$^{-1}$ for J, H and K$_s$ band, respectively.
The color and structure are in
sharp contrast  between the northeastern and southern parts. While the
northeastern rim is dominated by J and H band emission 
(represented with blue and green colors in Figure 1), the
southern ridge is dominated by K$_s$ band emission (red in Fig. 1).
The K$_s$ band emission defining  the southern
ridge is remarkably the same as those  H$_2$ maps (2.2$\mu$m) in Burton
et al. (1988) and Richter et al. (1995) except the 2MASS mosaic covers 
a larger field.
The J and H emission 
represents the first near-infrared imaging of IC 443.
Figures 3 (a)-(c) show the mosaiced
image after removing stars using  DAOPHOT.  Removal of the stars
was necessary to estimate the surface brightnesses shown in Table 1 
because a small contribution from 
faint stars would easily cause overestimation of the remnant flux.

We divided IC 443 into two global fields -- the northeastern rim  and
the southern ridge -- in order to compare and contrast the detailed
line diagnostics and physical parameters between fields. We have 
measured the surface brightnesses of the images  for a number of
positions,  verifying that the apparent color contrast between the 
northeast and southern fields is real (see Table 1). For the
northeastern position  we used ``Position 1" (R.A. 06$^{\rm 
h}$17$^{\rm m}$34.41$^{\rm s}$ and Dec. +22$^{\circ}$ 52$^{\prime}$
55.2$^{\prime\prime}$ epoch J2000)
where the intensities of [Fe~II] lines were measured by Graham et al. (1987).
Here the
targeted region  is defined by an ellipse with a radius of
48$''$$\times$18$''$ and a position angle of 300$^{\circ}$ (East
of North). 
The
southern position (R.A. 06$^{\rm h}$16$^{\rm m}$42.40$^{\rm s}$, 
Dec. +22$^{\circ}$ 32$^{\prime}$ 4.0$^{\prime\prime}$)
is where ISOCAM observations were carried out
(clump G of Cesarsky et al. 1999).
The ``Position 1"  and ``South (clump G)" positions are marked in Figure 3b. 
The ISOCAM region  is
37$''$$\times$69$''$ with an angle of 70$^{\circ}$.  Measured 
mean surface brightnesses 
are given in Table 1. The J:H:K$_s$ ratio represents the southern sinuous
ridge among the K$_s$ band dominated area (red in Figure 1). 
Among the K$_s$ band dominated area, the eastern side of IC 443 
shows different J:H:K$_s$ ratio from those in the southern sinuous ridge.
The southern sinuous ridge within which the  J:H:K$_s$ ratio
is similar to that of ``South (clump G)"
defines below Dec  +22$^{\circ}$ 30$^{\prime}$
00$^{\prime\prime}$ in the eastern side as well as the south.
The position ``East" in Table 1 (see Figure 3b)
shows brighter emission
in J and H (relative to K$_s$) when compared with that of the southern
sinuous ridge region. 
The estimated surface brightnesses in Table 1
have $\sim$30\% uncertainty
due to real surface brightness
variation within the elliptical regions.

We have measured J, H, and K$_s$ band surface brightnesses
from a few extra positions.
Within the northeastern rim (blue and green in Figure 1) 
the J:H:K$_s$ brightness ratios are similar to that of ``Position 1"
with less than 10\% variation. 
However, the K$_s$ dominated area (red in Figure 1) 
is subdivided into two area. 
One is the southern sinuous ridge and
the other is the eastern part of the remnant.
Within the southern sinuous ridge (including clump G position in the
southwest), the J:H:K$_s$ brightness ratios are also similar to that of 
``South" (clump G) position
with less than 10\% variation. 
However, 
in the eastern part,
the J:H:K$_s$ ratio gradually changes position by position from the
the ratio of
southern ridge to the ratio of northeastern rim.

The possible emission sources that contribute to the 2MASS near-infrared
emission 
include hydrogen recombination (P$\beta$, Br10 and Br
$\gamma$), molecular hydrogen lines (e.g. 2-0 S(1), 1-0 S(7) and 1-0
S(1)) and forbidden ionic lines (e.g. [Fe~II]). We assume there is negligible 
continuum within the 2MASS bands because the ISO observations show no
continuum for 5-14$\mu$m. The noise level of
the ISOCAM observation was 0.2$\times$$10^{-4}$  erg s$^{-1}$ 
cm$^{-2}$ sr$^{-1}$ which is comparable to the errors of
2MASS measurements in Table 1, and the near-infrared continuum brightness is
expected to be smaller than that of mid-infrared from 
heated small grains if they survived the shock 
(Dwek \& Arendt 1992). If there is continuum, it
would be less than 10\% of the 2MASS surface
brightnesses. 
Moreover,  previously measured line fluxes of
[Fe II] and H$_2$ were comparable to those of J, H and K bands
within errors (see detail in section 3 and 4).
The estimated non-thermal continuum 
due to synchrotron radiation, likewise, is too small to contribute 
to the 2MASS observed brightness. 

\subsection{ISO}

ISO LWS observations of [O~I] (63.2$\mu$m) were carried out for 11
positions of  the northeast field, crossing the northeastern rim with
an interval of 160 arcsec (illustrated in Figure 3b on the H band
image).  The beam size is 80$''$.  The observations took place 1998
Feburary 27 UT. The total time for these observations was 19 minutes,
with  approximately 1 minute dedicated to each spectrum. We averaged
the individual scans and fit the detected lines with gaussians. The
final signal-to-noise was satisfactory ($>>$ 10), clearly detecting the
targeted line from 7 positions which fall within the remnant. 
The spectra are shown in Figure 4 and the surface brightnesses
are listed in Table 2. 
The maximum brightness of the [O~I]
line at position ``e" 
is 4.8$\times 10^{-4}$ erg s$^{-1}$ cm$^{-2}$ sr$^{-1}$. The [O~I]
line is comparable to that of [Si~II] (34.8$\mu$m; Oliva et al. 1999a),
and is twice   as bright as [Ne~II] (12.8$\mu$m) after correction for
interstellar extinction.
The [O~I] line
emission peaks at the northeastern shell, where the 2MASS images showed
filamentary structure in J and H.  We measured 2MASS H band surface
brightnesses from the 11 positions where [O~I] brightnesses were
measured with the same beam, but large uncertainties
of 2MASS surface brightnesses over limited angular scales
hinder further comparison of the 
emitting region structures.
The background measurement error was large for relatively weak emission,
since the 80$''$ beam diluted the detected 2MASS emission.
The only significant detection 
is at the peak position ``e" 2.1$\pm$0.46 $\times 10^{-4}$ 
erg s$^{-1}$ cm$^{-2}$ sr$^{-1}$.
When we normalize the peak surface brightness between [O~I] and 
H band, the other expected detection in 2MASS
band images is   
from the position ``f", which would be 40\% of the peak surface and should
have been detected from 2MASS H band accounting for the errors. This
suggests possible different shock structures between 
[O~I] and [Fe~II]. However, with broad H band imaging, this is 
only a possibility.

\subsection{ CO observation}

We observed a small ($7^\prime \times 12^\prime$) region in the
northeastern interior portion of the remnant, centered on 
 R.A. 06$^h$17$^m$48$^{\prime\prime}$ and 
 Dec. 22$^\circ$42$^\prime$00$^{\prime\prime}$
(J2000) using the National Radio Astronomy Observatory 12-meter
telescope on Kitt Peak. The position is marked
in Figure 3c. The 1mm receiver was tuned to the 
CO(2$\rightarrow$1) line, and the spectra were processed with
the millimeter autocorrelator for single-point spectra
(resolution 0.06 km~s$^{-1}$) and filter banks
for mapping (resolution 0.65 km~s$^{-1}$). All spectra
were obtained with respect to an absolute reference position
located at  R.A. 06$^{\rm h}$18$^{\rm m}$30.0$^{\rm s}$
and Dec. +22$^{\circ}$ 54$^{\prime}$ 00.0$^{\prime\prime}$ (J2000).
Maps (see Figure 8) were obtained using the on-the-fly technique (Mangum 1997),
which yields well-sampled maps ($10^{\prime\prime}$)
with very uniform calibration.

\section{ Line Emission from the Northeastern Rim and Comparison with 
Shock Models}

The northeastern shell shows sheet-like filamentary structure similar
to the optical emission (Fesen \& Kirshner 1980; Mufson et al.
1986).  Emission in the J and H bands is equivalently bright for
the northeastern shell. The observed
J: H: K$_s$ surface brightness ratio is 1:0.7:0.1. 
This ratio is 1:0.5:0.08 after 
correcting for extinction, using 
A$_V$ = 3.1 mag (Fesen \& Kirshner 1980) and the extinction law 
given by Rieke \& Lebofsky (1985).
The 2MASS colors suggest very weak K$_s$
emission in the northeastern rim, $\sim$10\% in comparison
to the southern part. 
Previous observations indicate no  evidence of H$_2$ line emission
toward the northeastern rim (at the ``Position 1"; Graham et al. 1987). On the
other hand, K$_s$ band emission for the northeastern rim can be
explained by the Br$\gamma$ line alone. 

The possible sources for the
northeastern rim emission observed in the J and H bands are H recombination  and
[Fe~II] lines, discounting  H$_2$ emission based on its non-detection 
by Graham et al. (1987). First, we examined H recombination lines within
the 2MASS bands: J, H, K$_s$ bands cover P$\beta$ (1.28$\mu$m),  Br10
(1.74$\mu$m) to Br$\gamma$ (2.17$\mu$m), respectively. The Case B 
ratio of J: H : K$_s$ bands is 1:0.06:0.2 with 3\% changes depending on
the temperature (2000 - 20000 K). We ignore a few other hydrogen lines, 
because they are orders of magnitude weaker 
than Br10. 
The detailed line intensities 
relative to H$\beta$ depends on the temperatures as shown in Table
3. 

Second, we simulated the [Fe~II] line intensities within the 2MASS
bands. The [Fe~II] 1.64$\mu$m line has been shown to be strong in a
position on the northeast rim (Graham et al. 1987), 
with a surface brightness
of 1.9$\times 10^{-4}$ erg s$^{-1}$ cm$^{-2}$ sr$^{-1}$.
The Br$\gamma$ line surface brightness is 
0.04$\times 10^{-4}$ erg s$^{-1}$ cm$^{-2}$ sr$^{-1}$ using 
the  19.6$''$ aperture, which was much weaker than that of [Fe~II] 1.64$\mu$m.  
The H band surface brightness at ``Position 1,''
2.4$\times 10^{-4}$ erg s$^{-1}$ cm$^{-2}$ sr$^{-1}$,
is slightly brighter than the sum of two H-band lines
that have been measured spectroscopically,
[Fe~II] 1.64 $\mu$m (Graham et al. 1987) and Br~10 (see Table 4),
which is 1.91$\times 10^{-4}$ erg s$^{-1}$ cm$^{-2}$ sr$^{-1}$.
We have calculated line intensities of [Fe~II] among 13 levels with 45
total transitions by solving the excitation rate equations as a matrix
using the atomic data from Nussbaumer \& Storey (1988).  Our goal is 
to simulate approximately the lines within the 2MASS bands and examine
if the surface brightnesses are comparable to the observed ones.
We noticed that the
J:H ratio of [Fe~II] lines is largely based on the line ratio of
[FeII](1.25$\mu$m)/[FeII](1.64$\mu$m)=1.35, set entirely by the atomic
data due to both  lines having the same upper energy level
(see Nussbaumer \& Storey
1988). Figure 5
illustrates simulated J, H and K$_s$ band [Fe~II] lines, including 3 
lines in J, 5 lines in H (there were 6 lines each for J and H bands,
but weak lines are not shown), and no lines in K$_s$. The simulations show
that the J and H band emission from [Fe~II] largely depends on 
the 1.25$\mu$m and 1.64$\mu$m lines, respectively, because
those two lines are much brighter than 
others.     
The J:H ratio changes less than
$\sim$30 \%  from 1.35 when including other [Fe~II] lines
depending on the density and temperature.  When we used the 
[Fe~II] line ratio
between 17.93$\mu$m and 25.98$\mu$m 
of 0.5  and between [Fe~II] at 1.64$\mu$m and 17.9$\mu$m
of 4.2 (Graham et al. 1991; Oliva et al. 1999a),
we found a solution with 
a electron density of 500 cm$^{-3}$ and temperature of 12000 K.
Although these values have a large uncertainty due to different beam size
and possibly different filling factors,
it should be sufficient to estimate the contributions of 
[Fe II] lines to J and H bands.
Using these physical parameters, the sum of
[Fe~II] line brightnesses falling in J band, 4.0$\times 10^{-4}$ erg s$^{-1}$,
is 1.6 times as large as the sum of
[Fe~II] line brightnesses in H band, 2.5 $\times 10^{-4}$ erg s$^{-1}$ 
cm$^{-2}$ sr$^{-1}$. 
Using the
predicted estimation of [Fe~II] and H recombination lines, the total J, H and
K$_s$ surface brightness are 4.2$\times 10^{-4}$, 2.51$\times 10^{-4}$,
and  0.04$\times 10^{-4}$ erg s$^{-1}$ cm$^{-2}$ sr$^{-1}$,
respectively. 
These results are consistent with the observed surface
brightnesses (see Table 1) considering
various uncertainties such as the beam size and error of the 
each measurement. We
conclude that along the northeastern rim
the J band emission is composed of [Fe~II] ($\sim$90\%)
and P$\beta$ ($\sim$10\%), and the H band is  [Fe~II]  ($\sim$99\%) and Br10
($\sim$1\%), and the K band is Br$\gamma$.

We have listed the surface brightnesses of  [Fe~II] lines, [O~I] and
other detected infrared and optical lines from 
``Position 1" in the northeastern rim
(Oliva et al. 1999a; Fesen \& Kirshner 1980)  in Table 4. We have
compared these surface brightnesses  with various shock models. A fast
J shock model in Hollenbach \& McKee (1989; HM hereafter)  with a
density of 10$^3$ cm$^{-3}$ and a shock velocity of 80 - 100 km s$^{-1}$ is
consistent with most of the detected lines such as [Ne~II] (12.8$\mu$m),
[Fe~II] (17.93$\mu$m and 25.98$\mu$m) and [O~I] (63$\mu$m).
The comparison with HM89 model shows that  
a lower shock velocity with a density of 10$^3$ cm$^{-3}$ predicts  
much weaker [Ne~II] than observed, and  
all observed lines
are too weak for a higher density of
10$^4$ - 10$^6$ cm$^{-3}$. 
However, there
are still two important differences between the data and
the most consistent shock model with a
density of 10$^3$ cm$^{-3}$ and a shock velocity of 80 - 100 
km s$^{-1}$ of HM.  First,
multi-ionized ion species like [Ne~III], [S~III] and [O~IV] are not 
included in the model,  thus it is not clear whether  a fast J shock
model with V$_s$ = 80-100 km $s^{-1}$  can explain the highly ionized
species or a higher shock velocity  is required. Second, the observed 
[Si~II] (34.8$\mu$m) line is at least a factor of 5 stronger than that in
the HM model. We discuss below if these differences infer other types of
shocks, V$_s$, and/or n$_o$.

We compared other shock models
(Hartigan et al. 1987; McKee et al. 1984) using  shock diagnostic plots
for different
shock velocities and/or densities.
Since multi-ionized species are expected from high
velocity shocks, we have checked if the observed [Ne~III] requires a
higher shock velocity, or if it can be produced by V$_s$ $\sim$ 100 km
s$^{-1}$.  
The ratio of [Ne~III] to [Ne~II] is
shown as a function of shock velocity in Figure 6a
using the shock models of Hartigan et al. (1987). 
The observed ratio of
0.55$\pm$0.15 implies two possible shock velocities of V$_s$ = 100$\pm$20
km s$^{-1}$
and 250$\pm$30 km s$^{-1}$.  While the overall tendency is that the ratio
increases as the shock velocity increases, the ratio also shows a bump
around 100 km s$^{-1}$ which is due to the shape of the cooling function.
Figure 6a shows that the observed ratio of [Ne~III] and [Ne~II]  is
consistent with the J-shock models of V$_s$ $\sim$ 80-120 km s$^{-1}$.
The absolute surface brightnesses of [Ne~II],  [Fe~II] (25.98$\mu$m), 
and [Si~II] can
be used to measure the shock velocity: they are too weak for a high
velocity shock of $\sim$250 km s$^{-1}$ as shown in Figure 6b. The shock
velocity inferred from the brightnesses of three lines are marked with
a thick solid line in Figure 6b, where the allowed shock velocity, using the
overlapped shock velocity from the three lines, ranges between 80-120 km
s$^{-1}$. The surface brightness in Table 4 assumes that emission is
filled within the beam. So if the filling factor is in fact smaller, the
surface brightness is conversely larger than the values given. 
The slightly lower shock velocities 
are inferred even if 
[Ne~II] and [Fe~II] (25.98$\mu$m) have
a factor of 5 - 7 smaller filling factor. 
The
brightness of [Si~II] shows that if the surface brightness
(due to smaller filling factor) is greater than 1.5 times
of the current value in Figure 6b, no shock velocity solution is found. 
In summary, both the observed line brightnesses of [Ne~II], [Si~II] and
[Fe~II] and the ratio of [Ne~III]/[Ne~II] are consistent with V$_s$ of
$\sim$100 km s$^{-1}$,  and the line brightnesses are more than a factor of 4-5
lower than predictions for higher ($>$ 150 km s$^{-1}$) velocity shocks. 

 We have compared the detected infrared line intensities with those 
 for the remnant
RCW 103, which is undergoing a high velocity shock, V$_s$ $\sim$ 250  km
s$^{-1}$ (Oliva et al. 1999b). In contrast, many lines from 
IC 443 are much weaker
(at least an order of magnitude) than those of RCW 103, in particular
[Fe~II]. The intercloud shock velocity for IC 443 and RCW 103 is 
800  km
s$^{-1}$ (Petre et al. 1988) and 
12,000  km s$^{-1}$  (Nugent et al. 1984),  respectively, which 
is primarily related to
the age difference between remnants: IC 443 age is 5000 - 10000 yr
(Petre et al. 1988) and RCW 103 is 1000 yr (Nugent et al. 1984).
The shock velocity ratio between cloud medium and
and intercloud medium is inversely proportional to
the square root of the density ratio. 
The lower shock
velocity of 100 km s$^{-3}$ in IC 443 
compared with that of RCW 103 is probably related to the remnant age,
since the pre-shock density of RCW 103 (10$^3$ cm$^{-3}$)
is comparable to that of northeastern rim of IC 443. 
On the other hand, a high velocity shock greater than 150
km s$^{-1}$ is not stable; hence, comparison with the
steady  shock model may  not be meaningful for this scenario  (see
review in Draine \& McKee 1993; Innes et al. 1987).

We explored the possible pre-shock density range of the pre-shock medium at  the
northeastern rim.  Figure 6c shows a set of predicted line brightnesses
as a function of density based on McKee et al. (1984) where a density
ranges from 10 - 10$^6$ cm$^{-3}$, covering a larger density range
than  those in Hollenbach \& McKee (1989).  The observed surface
brightness for these lines at a    density of 300 - 1000 cm$^{-3}$ 
agrees with the  models in McKee et al. (1984) and Hollenbach \&
McKee (1989), except   for  the [Si~II] line, as shown in Figure 6c. 
Note however, the observed surface brightness of [Si~II] is  consistent
with those in Hartigan model (1991), as discussed before.  The
difference is due to metal abundance since the Hartigan models used
higher abundances  of Si  (3.23$\times 10^{-5}$) while the HM model 
used a smaller value  of 3.6$\times 10^{-6}$. While the Hartigan model
did not use dusty radiative transfer and assumed complete grain
destruction, the HM model includes dusty radiative transfer
and accounts for grain destruction. 
Therefore, it suggests that
either the intrinsic abundance of Si is higher than predicted or 
more efficient grain
destruction occurs. The grain destruction depends on a wide
array of physical parameters (detailed in Jones et al. 1996), requiring
further theoretical exploration.  Finally, a low pre-shock density of $\sim$10
cm$^{-3}$ for the near-infrared emitting region can be ruled out 
based on the observed
values for the northeastern rim of IC 443 as shown in Figure 6c; 
the observed brightnesses of [O~I] and [Fe~II] are too strong for
a density of 1-10 cm$^{-3}$. This conclusion is even true
if the filling factor of the emission in our beam is small.
The Cygnus Loop is consistent
with the shock model prediction of a density of 1-10 cm$^{-3}$ and  
a shock velocity of 100 km s$^{-1}$ (McKee et al. 1987). The model predicted [O~I]
and [Si~II] brightnesses 
for a density of 10 cm$^{-3}$ and a shock velocity of 30 km s$^{-1}$ are 
2$\times$10$^{-6}$ and 6$\times$10$^{-7}$ erg s$^{-1}$ cm$^{-2}$ sr$^{-1}$, 
respectively, which are 
too weak for the observed brightnesses of IC 443. A low density
($<$10 cm$^{-3}$) and a low shock
velocity model is not consistent with that of IC 443.  
We note there is a limitation in comparison of the observed brightnesses
with shock models, because the shock structures are not resolved
with available data. The observed values are measured 
with different beam sizes, and it is also possible that
the shock is tilted and/or  
multiple-shocks are present within the beam. 
Higher angular resolution imaging of the infrared lines 
are needed for further study.

\section{Shocked Molecular lines of the Southern Ridge and Comparison with
Shock models}

In contrast to the northeastern rim, the southern ridge is dominated by
K$_s$ band emission, exhibiting a knotty structure (Figure 3c). 
In this direction, the
extinction is 3.3 mag in A$_V$ (Burton et al. 1988), and the J:H:K$_s$
ratio of 0.2-0.33:0.36:1 becomes 0.35-0.5:0.4:1 after correcting for this extinction.
The shocked H$_2$ line emission, especially H$_2$ 1-0 S(1) (2.12$\mu$m) in K$_s$
band,  is well known from the southern ridge (Burton et al. 1988),
and it is produced  by an interaction with dense molecular clouds. 
The measured K$_s$ band surface brightness is 2.44$\pm$0.18 $\times$
10$^{-4}$ erg s$^{-1}$ cm$^{-2}$ sr$^{-1}$ (Table 1), which  
includes a dozen lines as shown in Figure 7b, 
among which the brightest lines are 1-0 S(1) (2.12$\mu$m), 1-0 S(2) (2.03$\mu$m),
and 1-0 S(0) (2.22$\mu$m).    
This brightness is approximately equal to 
a sum of the two detected lines 
(1-0 S(1) at 2.122$\mu$m, and 1-0 S(0) at 2.223$\mu$m) by
Burton et al. (1988) toward Clump C.
The predicted H recombination and H$_2$ line brightness and the
J:H:K$_s$ ratio using excitation 
temperatures ranging from 1000 - 4000 K, are
listed in Table 3. Note that the H$_2$ ratios are very sensitive to the
excitation temperature. 

Can the J and H band emission come from
hydrogen recombination lines$?$ Burton et al. (1988) has searched
spectroscopically for Br$\gamma$ lines, yielding a non-detection with
the upper limit of 0.042$\times$ 10$^{-4}$ erg s$^{-1}$
cm$^{-2}$sr$^{-1}$ (from 3$\times$10$^{-14}$ erg s$^{-1}$ cm$^{-2}$ for
19.6$''$ aperture diameter). Using the upper limit of the extinction
corrected brightness of
6$\times 10^{-6}$ erg s$^{-1}$ cm$^{-2}$ sr$^{-1}$, the P$\beta$ (for
J band) and Br10 (for H band) brightnesses should be 
less than 0.13$\times$10$^{-4}$,
and  0.01$\times$10$^{-4}$ erg s$^{-1}$ cm$^{-2}$sr$^{-1}$,
respectively.  It seems unlikely that the H-band flux is coming from the
hydrogen recombination line of Br10, although the flux of P$\beta$
might contribute to some portion of the J band flux. 
Using an upper limit for the Br$\gamma$ line, the P$\beta$
contribution could be as large as 7\% of  J band surface brightness.

We have estimated H$_2$ intensities within the 2MASS bands by deriving
excitation
temperatures from the many near- and mid-infrared spectral lines 
spanning the energy levels that contribute the lines in
the 2MASS bands.
The ISOCAM data from Cesarsky et al. (1999) for 5-14$\mu$m 
were fit to a two temperature model with  
the excitation temperatures T$_1$ = 657 K and T$_2$ = 1288 K
and the respective column density of 2.2$\times 10^{20}$,
and 0.22$\times 10^{20}$ cm$^{-2}$. 
The ISOCAM data (Cesarsky et al. 1999) and the ground-based 
near-infrared data by Richter et al. 
(1995) are shown in Figure 7a.
While the ISOCAM lines have upper energy levels  
of $<$7000 K, the ground-based data appear at 5000 - 26000 K. 
By fixing two excitation temperature components obtained from ISOCAM data,
a third temperature component (T$>$ 1288 K) is 
needed to fit both sets of data. 
Unfortunately, the ISOCAM data alone were not adequate to
constrain the high temperature component.
Therefore, we have developed a new two excitation temperature model to fit
both near-infrared and mid-infrared data.
The optimum values for the two 
temperatures and the corresponding column densities
are T$_{ex(1)}$$\sim$870 K and 
N$_1$$\sim$1.3$\times$10$^{21}$ cm$^{-2}$, and
T$_{ex(2)}$$\sim$2970 K and 
N$_2$$\sim$4.0$\times$10$^{18}$ cm$^{-2}$, respectively.
With this model, the estimated 
J:H band ratio is 0.8 and the H:K$_s$ band ratio is 0.2.
These ratios
are lower than observed values of 1.1 and 0.4, respectively.
We calculated the brightness of all lines that fall in the 2MASS bands.
The lines have upper energy levels of
11,000-14,000 K for J band, 10,000-19,000 K for H band,
and 12,000-15,000 K and 18,000-20,000 K for K band.
The contributed H$_2$ lines to 2MASS bands are shown
in Figure 7b.

There are two possibilities to explain the excess J and H band 
emission, relative to the H$_2$ model.
First, the excess J and H band emission might still  be 
from H$_2$ lines only,
considering that the accurate prediction requires non-steady C-shock
model as shown in Cesarsky et al. (1999). The fact that
the image morphology of J and H bands in the south,
in particular at the clump G position (see Figures 3a and 3c),
is very similar to that of K$_s$ band and different from
that of optical emission,
supports that the J and H band emission is from H$_2$ lines. 
It is possible that the non-steady C-shock model
could explain the observed J:H:K$_s$ ratio using only H$_2$
lines. When we increase the second temperature and decrease the
first temperature, we can obtain
a reasonable fit (except for the upper energy levels of 6000 - 8000 K), 
for which the H:K$_s$ ratio is close to 0.4.   
However, the J:H ratio greater than 1 can not easily be explained
using H$_2$ lines only.

The second possibility is that the remaining J and H band emission is  
from [Fe~II] lines, and some portion of J band emission is from P$\beta$.
We also measured another southern position 
(R.A. 06$^{\rm
h}$17$^{\rm m}$7.6$^{\rm s}$ and Dec. +22$^{\circ}$ 25$^{\prime}$
35$^{\prime\prime}$ epoch J2000; ``south ridge position") where an ISO SWS spectrum is available.  
The 2MASS J:H:K$_s$ ratio for this position
is similar to that of the position ``South (clump G)". 
ISO archival data show a clear detection
of [Si~II] (34.8$\mu$m), but the presence of [Fe~II] (26$\mu$m) is uncertain.
The upper limit of [Fe~II] (26$\mu$m) is 3$\times$10$^{-5}$ 
erg s$^{-1}$ cm$^{-2}$ sr$^{-1}$. 
A region with no [Ar II], [Ar III], and [Ne~II] (like in ISOCAM
data) can still emit 
[Si~II] and [Fe~II] because the latter has lower ionization
potentials.  
The brightness ratio between [Fe~II] 26$\mu$m and 1.25$\mu$m
is very sensitive to temperature; it can change between 0.2 -  
1000 with the higher temperature giving a lower ratio.   
A very hot, low-ionization gas could make enough emission to contribute
to the J and H bands at the levels observed; however, we lack sufficient
spectroscopic data to resolve the origin of the 
excess emission in the southern part of IC 443.

The far-infrared line [O~I] 63$\mu$m was first detected
in IC 443  by Burton et al. (1990). The authors suggested the presence
of two shocks to explain the [O~I] and the shorter-wavelength H$_2$
lines, either a C shock with a dissociative J shock with suppressed
oxygen chemistry, or alternatively, two C-shock models  (a high and low shock
velocity).  CO observations of Wang \& Scoville (1992) suggest the
presence of a J-shock from the broad line CO gas.
This implies that a mixture of J and C shocks are present in  a small scale
of clump.  However, recent  ISOCAM  observations toward  southern
region, clump G, shows that no ion lines were detected, while only
molecular hydrogen  lines  have been detected (Cesarsky et al. 1999).
A fast J-shock would have emitted rich lines from all abundant elements.
The ISOCAM observational result is rather surprising and remarkable given
that a J-shock component must be present (Burton et al. 1990). 
In comparison, other molecular interacting supernova remnants, 3C391,
W44, and W28 showed many atomic lines, including strong [Fe
II] and [Si~II], and molecular hydrogen lines with  ISO (Reach \&
Rho 2000).  Cesarsky et al. (1999) showed that the detected pure
rotation lines of molecular hydrogen  are consistent with a
time-dependent C-shock model with a shock velocity of 30 km s$^{-1}$
and a pre-shock density of  10$^4$cm$^{-3}$. Likewise, a C-shock with a
density of 10$^4$ cm$^{-3}$ and a shock velocity of 30 km s$^{-1}$ 
accounts for a large portion of the 2MASS band emission 
with H$_2$ lines. 
However, it is
still unclear if the broad CO line is due to a slow J-shock as suggested by
Wang \& Scoville (1994) or due to a C-shock.
Comparison of high resolution spectra of the two lines  would be an
important diagnostic for understanding  the type of shock for  [O~I] and
broad CO lines.

\section{Extent of the supernova molecular cloud interaction}

The shocked H$_2$ line map is well observed from the southern sinuous ridge
(Burton et al. 1988; Richter 1995). The shocked CO gas with broad lines
is also intensively observed from the same, southern region (Dickman
1988; van Dishoeck et al. 1995).  
The large field of view of the 2MASS image shows
that the H$_2$ emission extends to the east and the inner side of the
northeastern rim of the remnant as seen in Figure 1. The H$_2$
emission in the inner part of the northeastern rim is most clearly seen
in the mosaiced color image ($``$red" from Figure 1; the area is
marked with an arrow in Figure 3c). We measured the J:H:K$_s$ surface
brightnesses at
the position ($``$NE front" in Table 1: R.A.
06$^{\rm h}$17$^{\rm m}$48$^{\rm s}$ and Dec. +22$^{\circ}$
41$^{\prime}$ 58$''$) defined by an ellipse with a radius of
60$''$$\times$110$''$ and a position angle of 351$^{\circ}$. The K$_s$ emission is
much stronger than J or H, with the J:H:K$_s$ ratio of 
0.3:0.4:1 using Av=3.1 mag. This ratio is comparable to
that of the southern sinuous ridge,
suggesting  the H$_2$ lines are primary contributors. 
The molecular cloud, as traced with CO emission,
has been shown to stretch across the entire remnant,
which includes extending to the northeastern part 
(Cornett et al. 1977). 
The K$_s$ band emission from the inner part of the
northeastern rim suggests that this region is also  
interacting with the molecular clouds. The interaction
is likely the front side of IC 443, since the molecular   
clouds are in the front of the remnant (Cornett et al. 1977; Asaoka
\& Aschenbach 1994).

In order to determine whether the widely-distributed
K$_s$-band emission in the northern interior of the remnant 
is related to shocked molecular gas, 
we compare to CO observations.
A CO(1$\rightarrow$0) survey of IC~443 revealed numerous
shocked clumps, of which the smallest was labeled clump H (Dickman et al. 1992).
This clump is separated from the southern sinuous ridge, and it is
weaker than the others.
The 2MASS image in this region reveals many clumps embedded in 
filamentary emission. Our follow-up observations with the NRAO 12-m
have high enough angular resolution and sensitivity to allow a
detailed comparison. 
Figure 8 shows the CO map where K$_s$ 
emission is detected (where an arrow is marked
in Figure 3c) with three spectra.  The fact that broad CO lines are detected
where K$_s$ band emission appears, suggests that K$_s$
emission is due to  shocked H$_2$ lines.
The CO emission consists of a filament with a few distinct clumps. The
panels in Figure 8 show spectra at 3 positions along
the filament. The spectra consist of a broad component, with
a full-width-half-maximum (FWHM) of 14 km~s$^{-1}$ and a narrow emission
component with a FWHM of 1.2 km~s$^{-1}$.
The broad component of the CO emission is due to shocked gas; it does
not break into narrow components when viewed at high spatial or
spectral resolution. The width of the broad component is comparable
to other places in the southern ridge, and the velocity center follows
the systematic variation of the rest of the ridge (Dickman et al. 1992).
Our CO(2$\rightarrow$1) observations 
along with K$_s$ band image  therefore for the first time show that clump
H is part of the supernova-molecular cloud interaction.
The good correspondence between the CO map and the K$_s$-band map
in this region demonstrates that the physical properties of the 
shocked gas are probably similar to those in the southern 
sinuous ridge. By inference, then, it is almost certain that the
extended K$_s$-band emission at this location is due to H$_2$ line
emission.

The physical and shock properties between the northeastern rim
and southern ridge are compared in Table 5. Although the shock
parameters are primarily measured from $``$Position 1" and
$``$clump G" position, they are likely representing the  
 northeastern rim and southern sinuous ridge, respectively,
since 2MASS images
and surface brightnesses show similar J:H:K$_s$ ratios.
The strong H$_2$ lines are detected  for most of the southern
ridge while no ionic lines are detected so far from $``$clump G" position.
However, the J:H:K$_s$ ratio changes  gradually
from the south, to the east, and to the northeastern rim. 
The flux of the eastern position is
measured as shown in Table 1. 
The targeted region of  $``$East" position  is defined by an
ellipse with a radius of 36$''$$\times$71$''$ and an angle of
352$^{\circ}$ at the position of R.A. $6^{\rm h} 18^{\rm m} 15.5^{\rm s}$,
and Dec. $22^\circ$ 32$^{\prime} 11.4^{\prime \prime}$, J2000. 
The extinction A$_V$ = 3.17 mag, an intermediate value between the
northeastern rim and south, is used for $``$East" position.   
The J:H:K$_s$
ratios of the eastern positions correspond to those between the 
northeast and southern ridge. The
K$_s$:J ratio of 0.7 is much stronger than
that seen in the northeastern rim. 
This suggests that both [Fe~II] and H$_2$
emission can be  present from the eastern region.
One eastern region  
shows similar J:H:K$_s$ ratio at this position
and Graham et al. (1987) detected  both [Fe~II] and H$_2$ lines. 
Therefore, 
the emission lines of both ion and H$_2$  
must be combined in the eastern region, and
the regions of  J and C shocks could be associated.  
The C shock may be developed in a dense clump
while J shock is developed in a lower pre-shock density cloud surrounding  
the clump where the molecules are dissociated by a strong shock.
The eastern region shows combination of ionic and  H$_2$ lines 
as observed in other remnants including 3C391,
W44, and W28, within an ISO LWS beam (Reach \& Rho 2000). 
The clear distinction  between the northeastern and
southern sinuous ridge of IC 443 as inferred from
2MASS color is striking, delineating the  
J and C-shocks. 

\section{Near-infrared Luminosity of IC 443}

We estimated the total luminosity of J, H and K$_s$ bands, by summing
the emission in the starlight-subtracted image. The total flux is 11.6,
6.4 and 9.8$\times10^{-10}$ erg s$^{-1}$ cm$^{-2}$ for J, H, and K$_s$
band, respectively, with $\sim$20\% uncertainty due to the background
subtraction and residuals from removed bright stars.  
The extinction corrected luminosity using a distance to
IC 443 of 1.5 kpc is 6.5, 2.6 and 3.4  $\times$10$^{35}$ erg s$^{-1}$,
respectively. The total  near-infrared luminosity  is
1.3$\times$10$^{36}$ erg s$^{-1}$. 
The separate H$_2$, [Fe~II]  and hydrogen recombination line
luminosities within 2MASS bands are estimated based on the contributed
portion of their emission to J, H, and K$_s$ bands, respectively, which
was described in sections 3 and 4. We have first estimated
H$_2$ luminosity for J, H, and K band which is 1.3, 1.4, and 3.4
$\times$10$^{35}$ erg s$^{-1}$, respectively,  based on the H:K$_s$ of 0.4
and J:H of 0.8 and assuming that most of K$_s$ band is from H$_2$ lines.
The total H$_2$ luminosity of J, H and K$_s$ bands is 6.1$\times$10$^{35}$
erg s$^{-1}$.  The total [Fe~II] luminosity is 5.9$\times$10$^{35}$ erg
s$^{-1}$ based on 90\% of the J band and 99\% of H band flux after
excluding J and H band contribution from H$_2$ emission. The hydrogen
recombination line luminosity is 10\% of J band and 1\% of H band flux
after excluding J and H band contribution from H$_2$ emission. The
total hydrogen recombination line luminosity is 0.53 $\times$10$^{35}$
erg  s$^{-1}$. These luminosities have large uncertainties because 
fixed contribution ratios of H II, H$_2$ and [Fe II] (section 4 and 5)
are used from the northeastern rim 
and the southern ridge, respectively, 
and it did not account for the gradual changes in
the flux ratios to the 
east and to
northeast. In addition, there is still some uncertainty in  understanding
the J and H band emission in the south.  In summary, the total luminosities of 
H$_2$, [Fe~II]  and hydrogen recombination lines in the near-infrared bands are 
6.1$\times$10$^{35}$, 5.9$\times$10$^{35}$ and  0.53$\times$ 10$^{35}$
erg s$^{-1}$, respectively.

\section*{Conclusions}

1. Near-infrared emission from IC 443  was detected in J, H
and K$_s$ in the Two Micron All Sky Survey (2MASS) Atlas Images, showing a
shell-like morphology.  The northeastern shell is 
delineated into a sheet-like
filamentary structure with the J and H band emission dominanting the 
near-infrared, while
southern sinuous ridge emission shows knotty structure with K$_s$ band
emission dominant. These are the first J, H, K$_s$ near-infrared
images that cover the entire remnant. The color and structure show contrast
between the northeastern and southern parts as shown in Figure 1.

2.  The northeastern shell exhibits equivalent brightness between the J and H
bands, with a J:H:K$_s$ surface brightness ratio of 1:0.5:0.08 
after extinction correction.
The K$_s$ band emission is likely 
from Br$\gamma$ recombination, since
no H$_2$ lines have been previously detected.
We estimated possible [Fe II] lines and their line 
strengths with 2MASS bands by solving the excitation rate  equations
and using  the 2MASS fluxes and previously detected [Fe II] lines.
The primary line ratio of 1.24$\mu$m/1.64$\mu$m $\sim$
1.35 set by atomic physics was mainly important to determine
the J:H ratio, and
the total [Fe~II] line brightness falling in J band is 1.6 times as large as
[Fe~II] line brightness in H band estimated from the observed line ratios.  
Thus we conclude that the H band emission is
composed of [Fe~II] ($>$99\%) and Br10 ($<$1\%), and the J band emission of
[Fe~II] ($\sim$90\%) and P$\beta$ ($\sim$10\%).

3. We obtained positive detections of [O~I] (63$\mu$m) for 11
positions in the northeast using ISO LWS observations.
The [O~I] line
emission peaks at the northeastern shell, where the 2MASS images showed
filamentary structure in J and H.
The surface brightness of the brightest [O~I] line
corresponds  to a surface brightness of  4.8$\times
10^{-4}$ erg s$^{-1}$ cm$^{-2}$ sr$^{-1}$, which is 
comparable to that of [Si~II].

4. In contrast to the northeastern rim, the southern ridge is dominated by
K$_s$ band emission, with the extinction-corrected J:H:K$_s$ 
ratio of 0.35-0.5:0.4:1.
The shock-tracing H$_2$ 1-0 S(1) 2.12$\mu$m line
is clearly mapped for the southern sinuous ridge.   
We have developed a new two excitation temperature model to fit
both near-infrared ground-based and mid-infrared ISOCAM data.
The derived value for 
the J:H band ratio is 0.8, and for the H:K$_s$ band ratio is 0.2, which
are smaller than those observed. Therefore, 
most of K$_s$ band and approximately half of J
and H band emission can be explained  
only with  the H$_2$ lines. Some contribution of P~$\beta$ 
and other ionic lines to J and H bands is possible.

5. Comparison of the dominant shocks and the physical parameters
between  the northeastern rim and  the southern  sinuous ridge is
summarized in Table 5. The infrared emission  from  the northeastern
rim is mostly from ionized atomic lines and  some
neutral oxygen line [O~I], with no evidence of molecular hydrogen lines.  A
fast J-shock model with a shock velocity of 80-110 km$^{-1}$ and a
density of 10$<$n$_o$$\simlt$1000 cm$^{-3}$  is consistent with  the
observed  lines from the northeastern  rim.
The infrared emission
from  the southern ridge comes from  molecular H$_2$ and [O~I].
The dominant shock is a slow C-shock with a v$_s$= 30 km
s$^{-1}$ and  n$_o$=10$^{4}$ cm$^{-3}$.

\section*{Acknowledgement}
This publication makes use of data products from the Two Micron All Sky
Survey, which is a joint project of the University of Massachusetts and
the Infrared Processing and Analysis Center, funded by the National
Aeronautics and Space Administration and the National Science
Foundation. It is also based on observations with the Infrared Space 
Observatory (ISO). ISO is an ESA project with instruments funded
ISO is an ESA project with instruments funding by ESA Members States
and with the participation of ISAS and NASA.
We would like to thank  Gene Kopan for making initial mosaiced
images of IC 443. J. R. thanks the 2MASS Principal Investigator, Michael
Skrutskie for his encouragement on this project, and Alberto
Noriega-Crespo for helpful discussion on shock models. 
We thank Pierre Cox for sharing the ISOCAM results prior to publication
and helpful discussion. We also thank the anonymous referee for  
valuable comments which helped us to significantly improve
this paper. We acknowledge
the support of the Jet Propulsion Laboratory, California Institute
of Technology, which is operated under contract with NASA.

\clearpage

\section*{Figures}

\figcaption{Mosaiced 2MASS Atlas Image of IC 443. J, H and K$_s$ images
show color contrast between northeastern rim (J and H emission shown
with  blue and green color, respectively) and southern ridge (K$_s$ emission
shown with red color).  The northeastern rim is dominated by [Fe~II] and P$\beta$
(J) and [Fe~II] (H), while southern sinuous ridge emission is mostly
H$_2$ lines. [FIGURE 1 CAN BE RETRIEVED AS A JPG FILE FROM ASTRO-PH]}

\figcaption{Surface brightness profile across the RA direction
after the images are mosaiced for (a) J, (b) H, and
(c) K$_s$ bands, respectively. Dotted lines are backgrounds at the northern
portion
and at the southern portion of the mosaiced image, and solid lines for J
and H are
crossing
Dec. $22^\circ$ 52$^{\prime} 55^{\prime \prime}$
(J2000), and for K$_s$, crossing
the position of
Dec. $22^\circ$ 25$^{\prime} 24^{\prime \prime}$
(J2000).}

\figcaption{J, H and K$_s$ images after stars are removed:
(a) J-band image, with surface brightness ranging
0.3-6.4$\times$10$^{-4}$ erg s$^{-1}$cm$^{-2}$sr$^{-1}$.
(b) H-band image, with brightness ranging from 0.2-4.0$\times$10$^{-4}$ erg
s$^{-1}$cm$^{-2}$sr$^{-1}$.
(c) K-band image, with brightness ranging from  0.1-12$\times$10$^{-4}$
erg s$^{-1}$cm$^{-2}$sr$^{-1}$.
In panel (b), the 11 positions of ISO LWS  beam for [O~I] observations, 
``NE front",
``Position  1", and clump G (south) positions 
are also marked.  [FIGURE 3 CAN BE RETRIEVED AS A JPG FILE FROM ASTRO-PH]}

\figcaption{
ISO LWS [O~I] Spectra of 11 positions, and the y-axis ranges are adjusted
to show weaker lines of the spectra on the right panel. 
The brightest line at (e) has a
flux density of 4.8$\times 10^{-4}$  erg cm$^{-2}$s$^{-1}$sr$^{-1}$.
Spectra from top to bottom are sequential from  outside to the interior of the
 remnant,
for the positions marked in Figure 3b.
The peak position coincides with northeastern filaments shown in the 2MASS
image.}

\figcaption{ 
Estimate of the contribution of [Fe~II] line intensity
to J, H, and K$_s$ band
emission. }

\figcaption{
(a) The [NeIII]/[Ne~II] ratio as a function of shock
velocity. The upper and lower limits from the errors are marked
as dotted lines. (b) Constraint on a shock velocity using the line
brightnesses of [Ne~II], [Fe~II] and [Si~II] lines. The allowed shock
velocity ranges are marked with thick lines for each line brightness.
c) Shock model predictions of the line brightnesses
as a function of density from McKee et al. (1984) 
using a shock velocity of 100 km s$^{-1}$.
The observed values are marked for [O~I], [Ne~II], [Fe~II]
and [Si~II] as an asterisk, diamond, square, and triangle, respectively.
}

\figcaption{(a) Population level diagram as a function of upper state
energy for H$_2$ lines. Near-infrared (Richter et al. 1995)
and mid-infrared ISOCAM data (Cesarsky et al. 1999) are shown with squares and
crosses, respectively, from which a two temperature
model (solid line) of excitation are derived. The two dotted lines are
T$_{ex(1)}$=870 K and T$_{ex(2)}$ = 2970 K. 
(b)  Contributed H$_2$ line intensity
estimation to J, H, and K$_s$ band emission. }

\figcaption{
CO(2$\rightarrow$1) map contours.
The map center is at R.A. 06$^{\rm h}$17$^{\rm m}$48$^{\rm s}$
and Dec. +22$^{\circ}$ 42$^{\prime}$ 00.0$^{\prime\prime}$ (J2000),
where an arrow is marked in Figure 3c.
On the
right-hand side are three CO(2$\rightarrow$1) spectra 
along the filament,
showing the separation into broad and narrow components.}

\clearpage
\begin{table}[h]
\label{tflux}
\caption{Measured in-band surface brightnesses using 2MASS images}
\begin{center}
\begin{tabular}{|l|l|l|l|l|l|}
\hline \hline
Band    &  J & H & K$_s$ & J:H:K$_s$ \\
\hline \hline
{Northeastern rim (Position 1) }\span\omit\span\omit\span\omit\span\omit \\
\hline
flux$^a$    & 2.28$\pm$0.30 &1.58$\pm$0.39 & 0.28$\pm$0.26 & 1:0.7:0.1 \\
unred$^b$ &  4.49$\pm$0.60 & 2.4$\pm$0.60 &  0.37$\pm$0.34 & 1:0.5:0.08 \\ 
\hline \hline
{NE front} \span\omit\span\omit\span\omit\span\omit \\
\hline
flux & 0.06$\pm$0.05 & 0.10$\pm$0.09 & 0.30$\pm$0.10 & 0.2:0.3:1 \\ 
unred &0.12$\pm$0.11 & 0.15$\pm$0.14 & 0.40$\pm$0.13 & 0.3:0.37:1\\ 
\hline \hline
{East} \span\omit\span\omit\span\omit\span\omit \\
\hline
 flux & 0.73$\pm$0.18 & 1.19$\pm$0.43 & 0.81$\pm$0.15 & 1:1.6:1.1 \\
unred & 1.62$\pm$0.4 & 1.95$\pm$0.71 & 1.12$\pm$0.21 & 1:1.2:0.7 \\
\hline \hline
{South (clump G)} \span\omit\span\omit\span\omit\span\omit \\
\hline
flux & 0.80$\pm$0.23 & 0.87$\pm$0.28 & 2.44$\pm$0.18 & 0.33:0.36:1 \\
unred & 1.88$\pm$0.54 &  1.48$\pm$0.48 & 3.44$\pm$0.25  & 0.5:0.4:1 \\
\hline\hline
\end{tabular}
\end{center}
\tablenotetext{a}{The surface brightnesses have units of 10$^{-4}$erg 
s$^{-1}$ cm$^{-2}$sr$^{-1}$.}
\tablenotetext{b}{
$``$unred" indicates the extinction corrected brightness.}
\end{table}

\begin{table}
\caption{Surface brightnesses of [O~I] lines with the positions using ISO}
\begin{center}
\begin{tabular}{lllll}
\hline \hline
position &     RA (J2000) &  DEC   &   surface brightness  &    \\
    &&        &      10$^{-4}$erg s$^{-1}$cm$^{-2}$sr$^{-1}$ &\\
\hline
a & 06:18:49.7 &  +22:51:18.5 &   0.142 ($\pm$0.018)  &   \\
b & 06:18:37.7 &  +22:50:08.6 &   0.202 ($\pm$0.023)  &  \\
c & 06:18:25.7 &  +22:48:58.7 &   0.135 ($\pm$0.034)  &  \\
d & 06:18:13.7 &  +22:47:48.8 &   0.120 ($\pm$0.054)  &  \\
e & 06:18:01.7 &  +22:46:38.8 &   4.814 ($\pm$0.150)  &  \\
f & 06:17:49.7 &  +22:45:28.7 &   2.213 ($\pm$0.024)  &  \\
g & 06:17:37.7 &  +22:44:18.6 &   0.756 ($\pm$0.044)  &  \\
h & 06:17:25.7 &  +22:43:08.4 &   0.653 ($\pm$0.023)  &  \\
i & 06:17:13.8 &  +22:41:58.2 &   0.692 ($\pm$0.031)  &   \\
j & 06:17:01.8 &  +22:40:47.9 &   0.350 ($\pm$0.023)  &   \\
k & 06:16:49.8 &  +22:39:37.5 &   0.403 ($\pm$0.020)  &   \\
\hline
\end{tabular}
\end{center}
\end{table}

\begin{table}
\label{h2preda}
\caption{Temperature dependent Hydrogen recombination and molecular H$_2$ line intensities }
\begin{center}
\begin{tabular}{|l|l|l|l|l|l}
\hline \hline
H Recombination lines (relative to H$\beta$)\span\omit \span\omit \span\omit \span\omit \\
\hline
T (K) & J (P$\beta$)  & H (Br10)  & K$_s$ (Br$\gamma$)  & J:H:K$_s$ \\
2500 K      & 0.21  & 0.0112  & 0.0387 & 1:0.05:0.18   \\
5000 K      & 0.187 & 0.0106  & 0.0332 & 1:0.057:0.18 \\ 
10000 K     & 0.165 & 0.0092 & 0.028  & 1:0.056:0.17  \\
20000 K        & 0.146 & 0.008 & 0.0237  & 1:0.055:0.16  \\
\hline\hline
H$_2$ Line Predictions \span\omit \span\omit \span\omit \span\omit \\
\hline
 T$_{ex}$ (K)       &   J         & H     & K$_s$ & J:H:K$_s$  \\
1000 K  &1.25$\times 10^{-5}$ &1.31 $\times 10^{-5}$ & 0.001073&0.011:0.012:1\\
2000 K  & 4.9$\times 10^{-3}$ & 5.8$\times 10^{-3}$ &0.034704 &0.14:0.167:1 &\\ 
3000 K       & 0.033 &0.041 &0.1165 & 0.28:0.35:1  \\
4000 K      &0.08117 & 0.09887 &0.2148& 0.38:0.46:1 \\ 
\hline
\end{tabular}
\end{center}
\end{table}

\singlespace
\begin{table}
\caption{IR Lines and their fluxes in the northeastern rim} 
\begin{center}
\begin{tabular}{llclcll}
\hline \hline
 Line &  $\lambda$ & unredden flux & $^a$beam & Surface Brightness & $^b$ref \\

     & ($\mu$m) &(10$^{-13}$ erg s$^{-1}$ cm$^{-2}$) & ($''$) & (10$^{-4}$ erg s$^{-1}$ cm$^{
-2}$ sr$^{-1}$) & \\           
\hline
 H$\beta$    & 4861\AA & 20.8 & 2.8$\times$40 &   8   & 1  \\  
 $[$Fe II$]$  & 1.257    &  -   &  model         & 2.6   & this work \\
 P$\beta$     &  1.28      &  -   &  model         & 0.2  & this work \\
 $[$Fe II$]$ in J band &  6 lines  & -& model &  4.0 & this work \\
  Total J band      &   -       & -    & -        & $^c$4.2, 4.49$\pm$0.63 & this work \\
$[$Fe II$]$  & 1.644  & 13.3   & 19.6          &  1.9   &  2\\
 $[$Fe II$]$ in H band &  6 lines  & -& model &  2.5 & this work \\
 Br10       &   1.74       &     -    & model    & 0.012 &  this work \\
 Total H band  &  -    & -       &   -   & $^c$2.51, 2.4$\pm0.6$ &  this work \\
 Br$\gamma$ & 2.165    & 0.3  & 19.6 & 0.04 & 2\\
 $[$Ne II$]$   & 12.8  & 18   & 14$\times$27   & 2.02$\pm$0.34 & 3  \\
 $[$Ne III$]$  & 15.56  & 10   & 14$\times$27   & 1.12$\pm$0.23 & 3 \\
 $[$Fe II$]$   & 17.93  & 4   & 14$\times$27   &0.45 & 3\\
 $[$S III$]$  & 18.71  & 4    & 14$\times$27   & 0.45 & 3\\
 $[$O IV$]$   & 25.88  & 3    & 14$\times$27   &0.34  & 3 \\
 $[$Fe II$]$   & 25.98  & 8   & 14$\times$27   & 0.89$\pm$0.11 & 3 \\
 $[$Si II$]$   & 34.8   & 52     & 20$\times$33    & 3.35$\pm$0.45 & 3 \\  
 $[$O I$]$     & 63     & 570    &   80         &   4.80  &this work\\
\hline \hline
\end{tabular}
\end{center}
\tablenotetext{a}{ A single number indicates the diameter of a circular beam.}
\tablenotetext{b}{Reference: (1) Fesen \& Kirshner 1980, (2) Graham et al. 1987
(3) Oliva et al. 1999a}
\tablenotetext{c}{The two numbers are  model 
 and observed surface brightnesses, respectively.}
\end{table}

\doublespace
\begin{table}
\caption{Comparison of physical properties
of infrared emitting regions between the northeastern rim and the south }
\begin{center}
\begin{tabular}{|l|c|c|c|c|}
\hline 
Parameters & Northeastern Rim         &  South Sinuous Ridge\\
\hline
 Dominant IR lines       & [Fe~II], [Ne~II],[Si~II], [O~I]&   H$_2$, [O~I]  \\
\hline
Type of Shock   &       fast J-shock  & C-shock \\ 
\hline
Density (n$_o$)   & 10$<n_o<10^3$ cm$^{-3}$  & $\sim$10$^{4}$ cm$^{-3}$ \\ 
\hline
Shock velocity (V$_s$) & 80-110 km s$^{-1}$   & 40 km s$^{-1}$ \\
\hline
\end{tabular}
\end{center}
\end{table}

\begin{figure}
\centerline{\psfig{figure=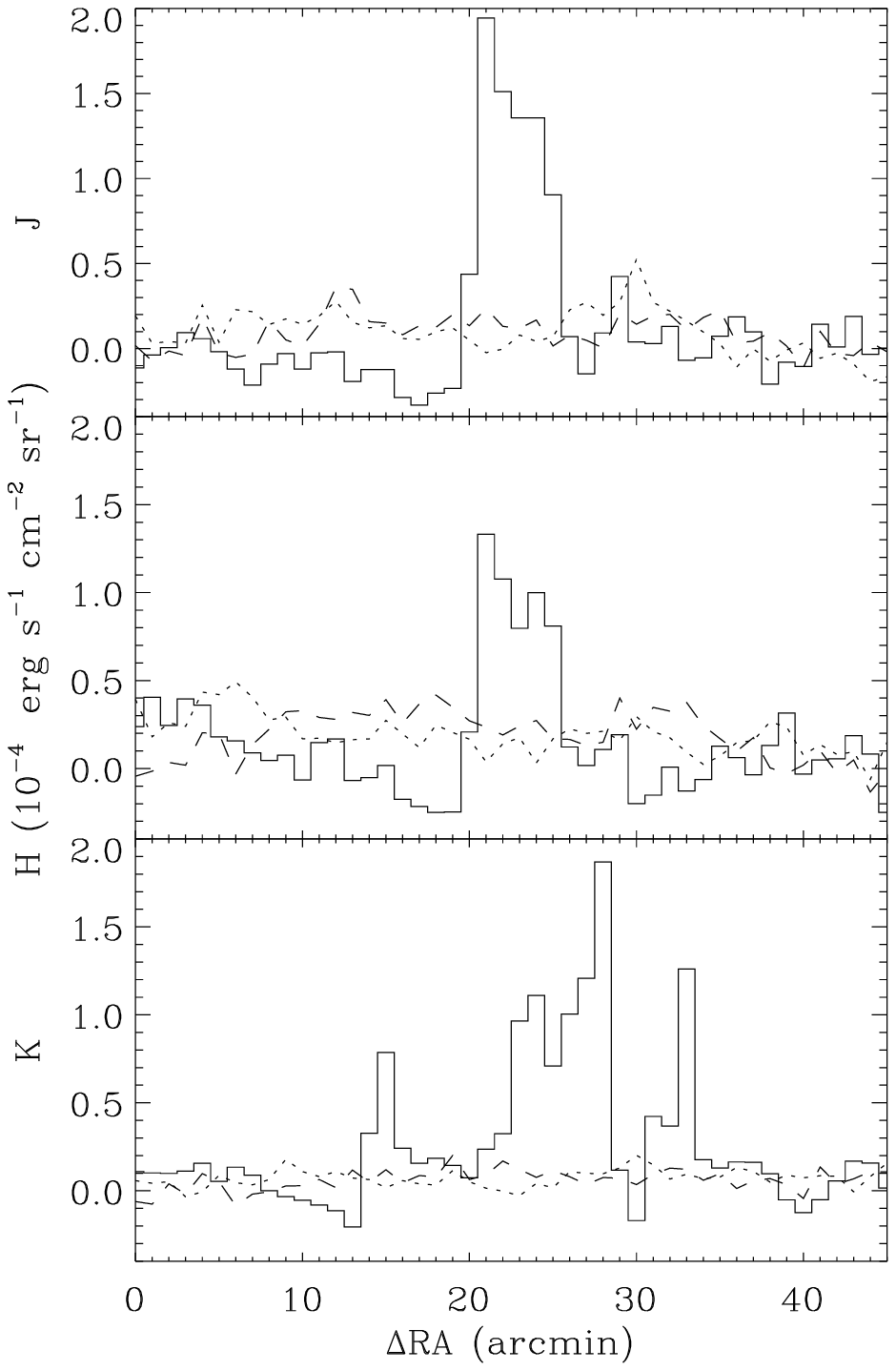,width=8cm}}
{Fig. 2}
\end{figure}

\def\extra{
\clearpage
\begin{figure}
\centerline{\psfig{figure=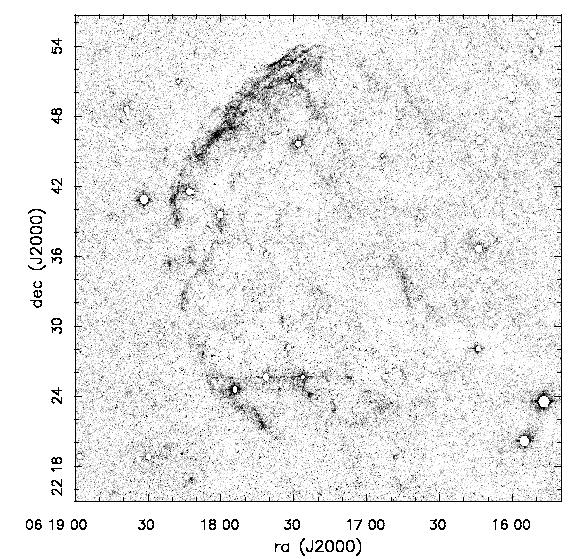,width=18cm}}

{Fig. 3a}
\end{figure}

\clearpage
\begin{figure}
\centerline{\psfig{figure=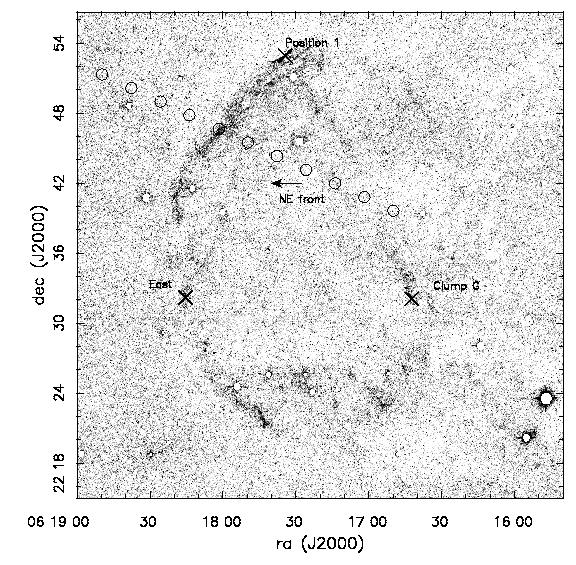,width=18cm}}

{Fig. 3b}
\end{figure}

\clearpage
\begin{figure}
\centerline{\psfig{figure=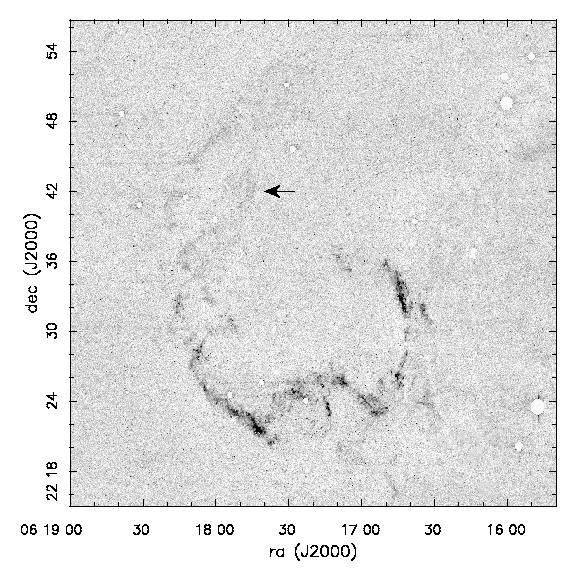,width=18cm}}

{Fig. 3c}
\end{figure}
}

\clearpage
\begin{figure}
\centerline{\psfig{figure=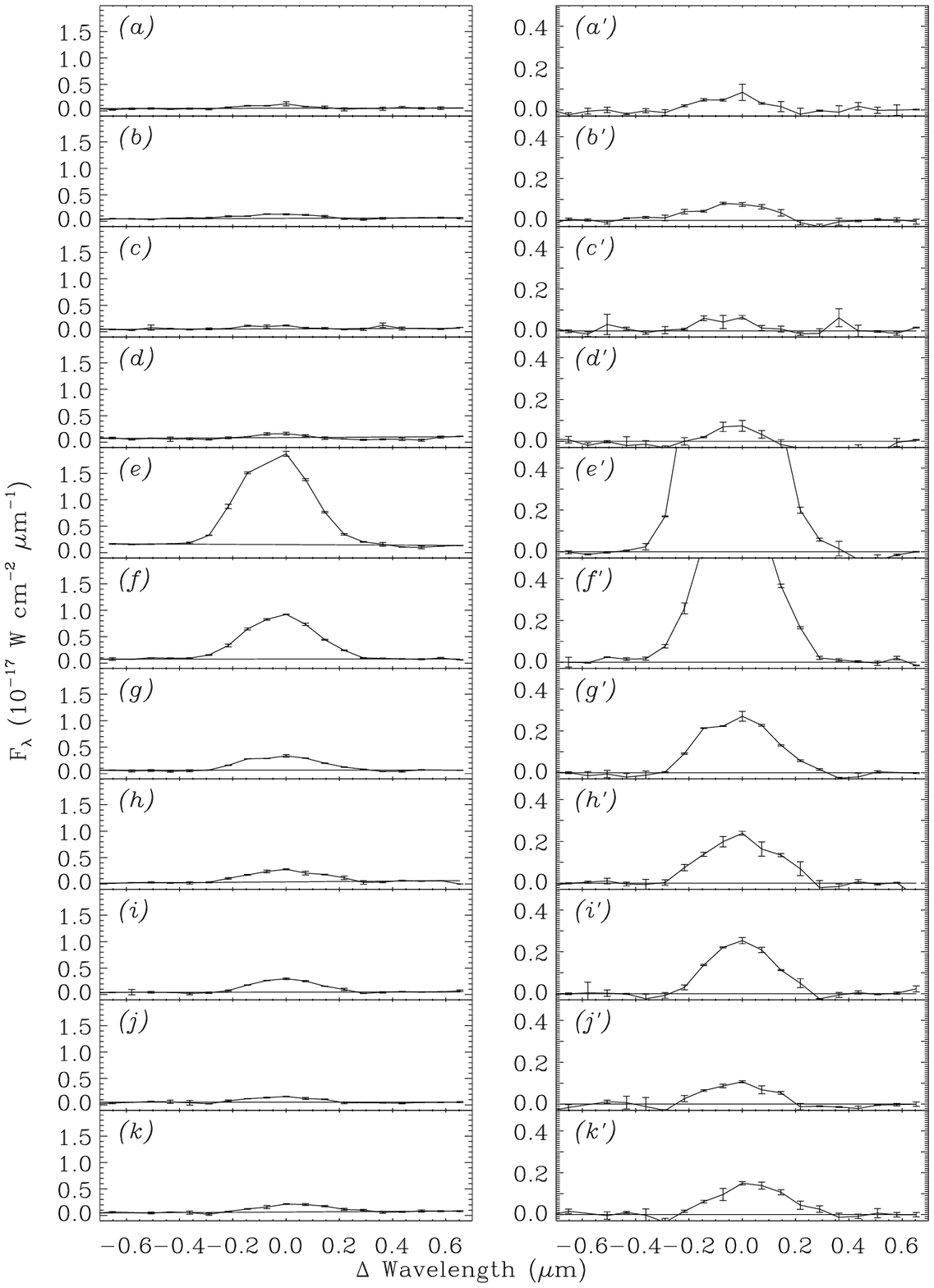,width=15cm}}

{Fig. 4}
\end{figure}

\clearpage

\begin{figure}

\centerline{\psfig{figure=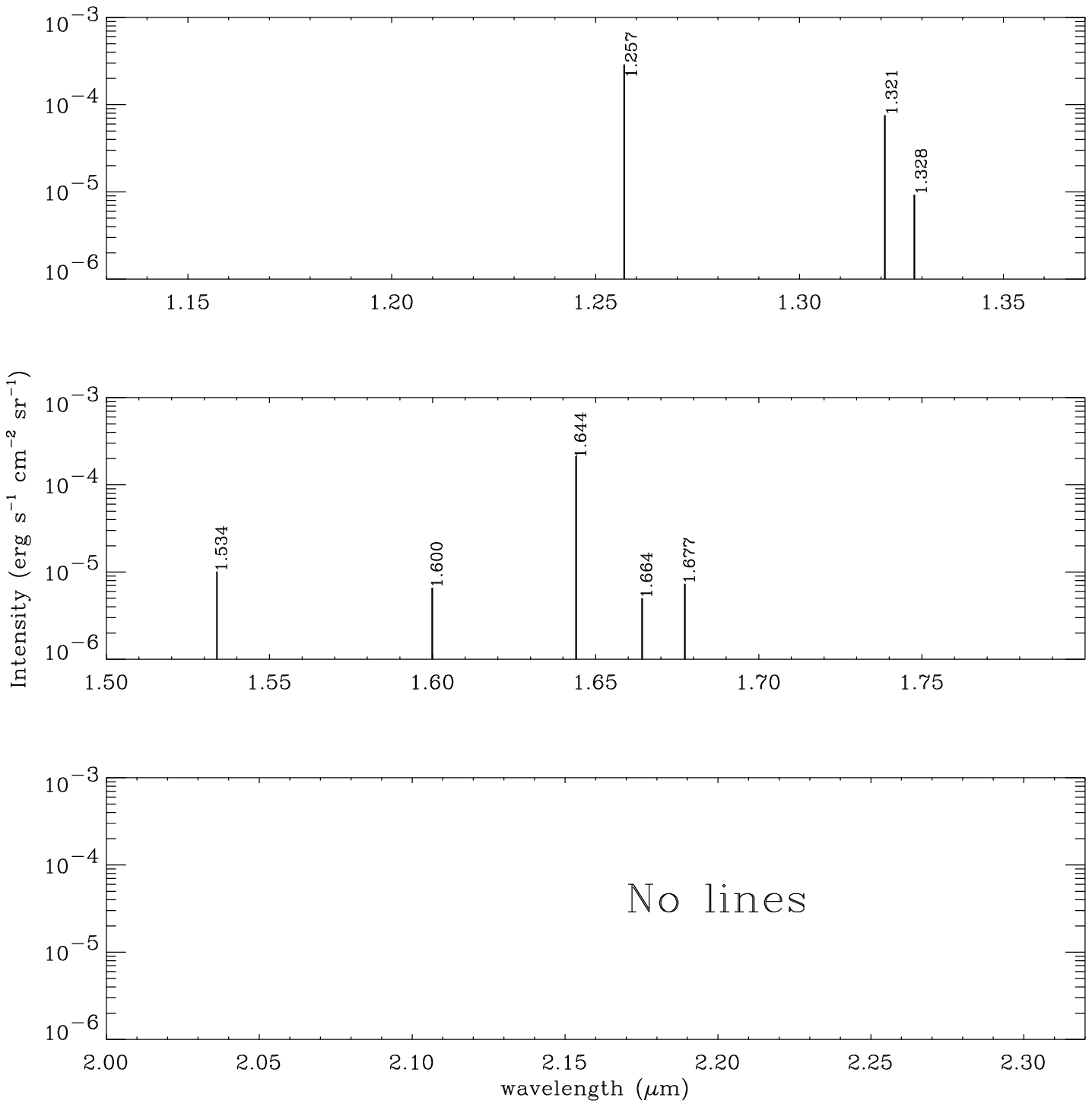,width=15cm}}

{Fig. 5}
\end{figure}

\clearpage
\begin{figure}
\psfig{figure=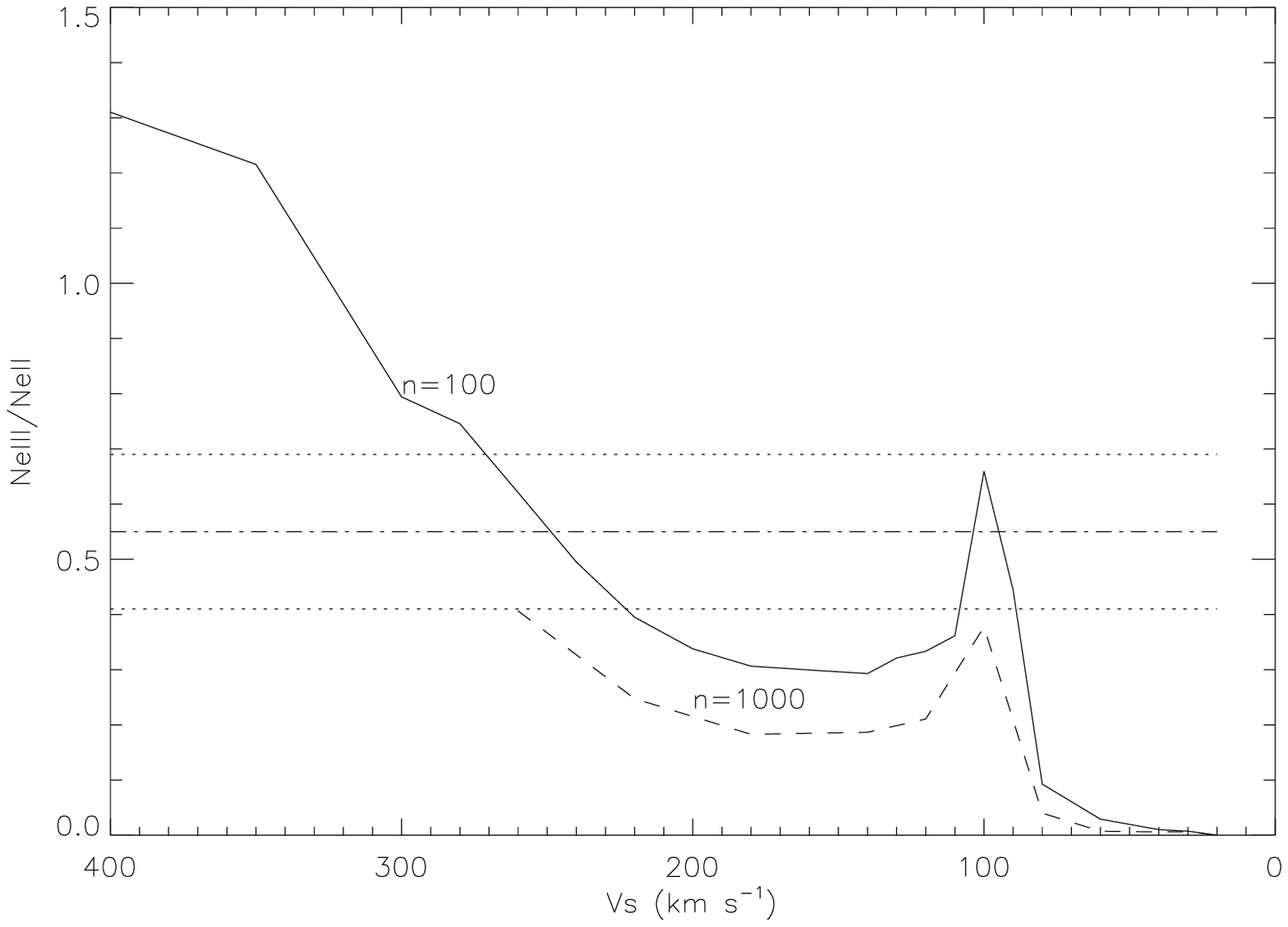,height=8truecm}

{Fig. 6a}
\end{figure}

\begin{figure}
\psfig{figure=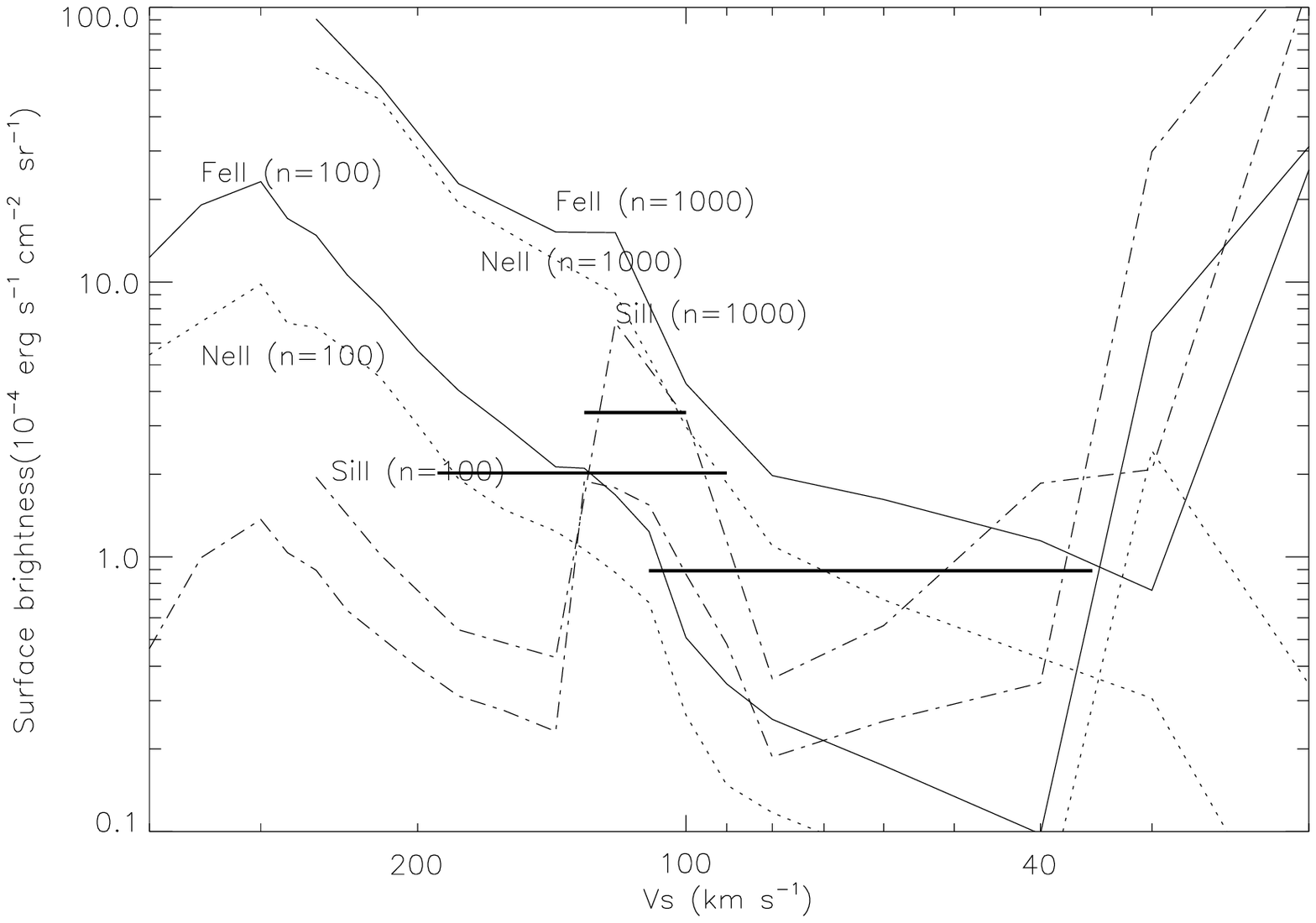,height=8truecm}

{Fig. 6b}
\end{figure}

\clearpage
\begin{figure}
\psfig{figure=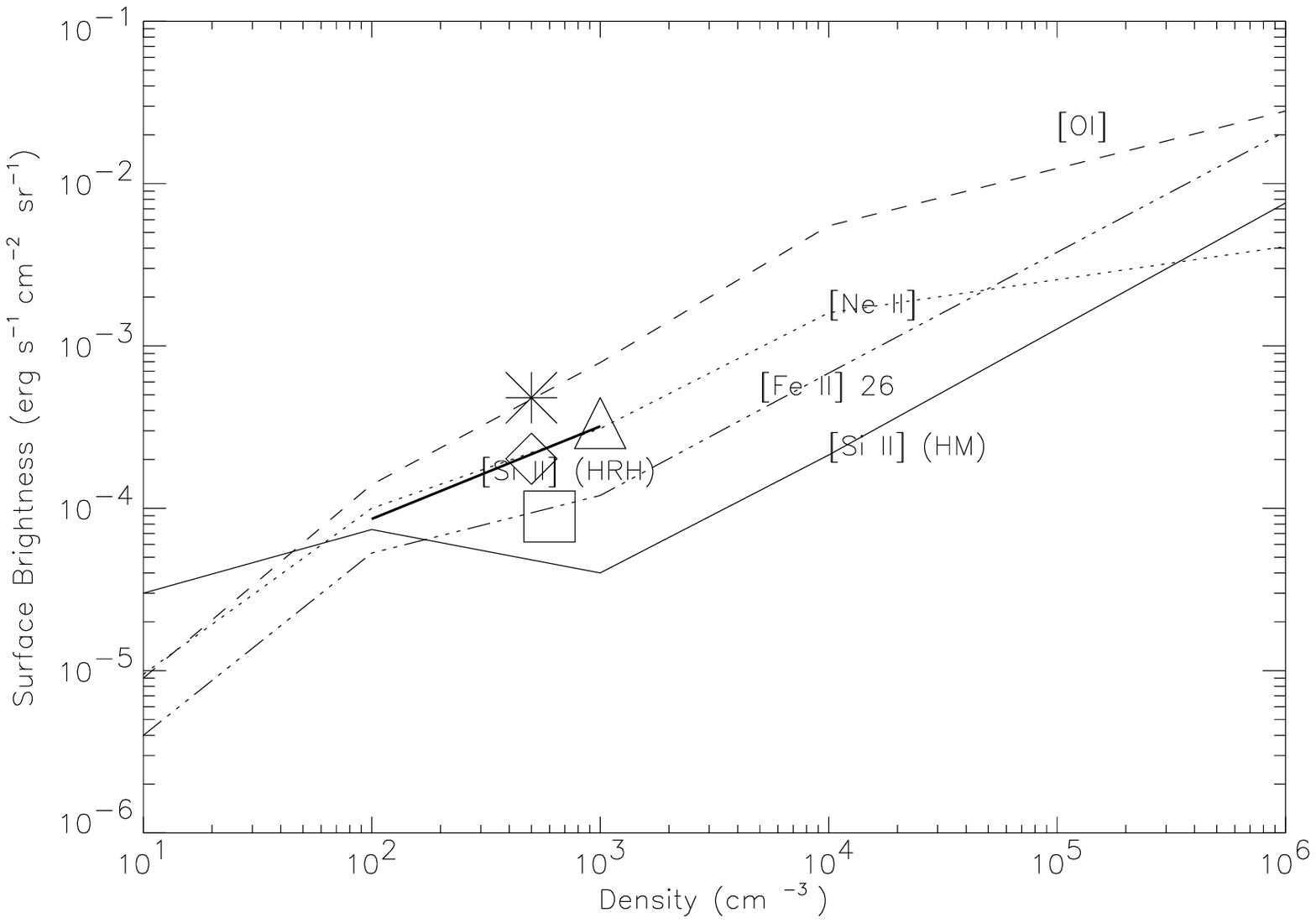,height=8truecm}

{Fig. 6c}
\end{figure}

\clearpage
\begin{figure}
\centerline{\psfig{figure=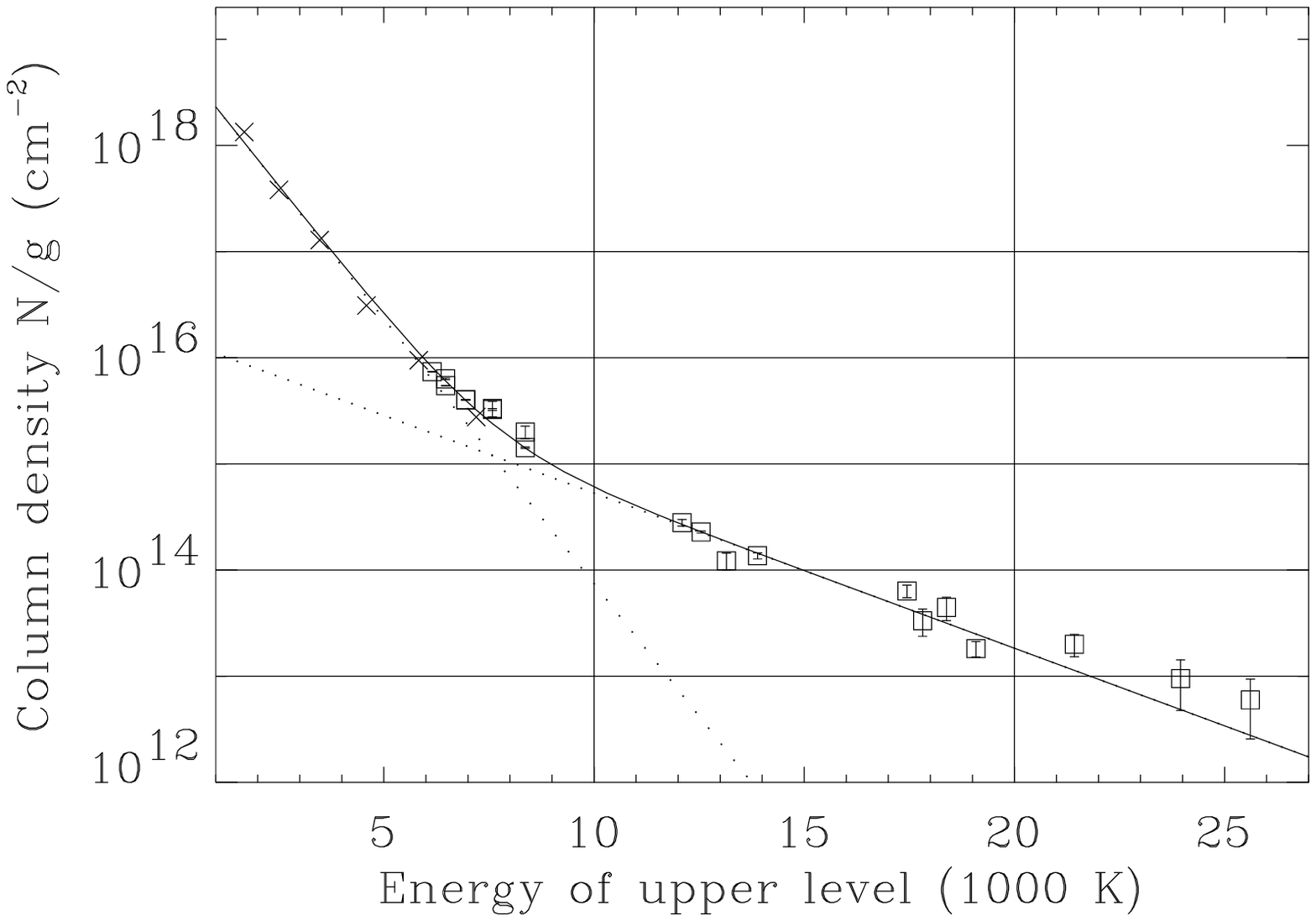,width=13cm}}

\vskip 1truecm
{Fig. 7a}

\end{figure}

\begin{figure}
\centerline{\psfig{figure=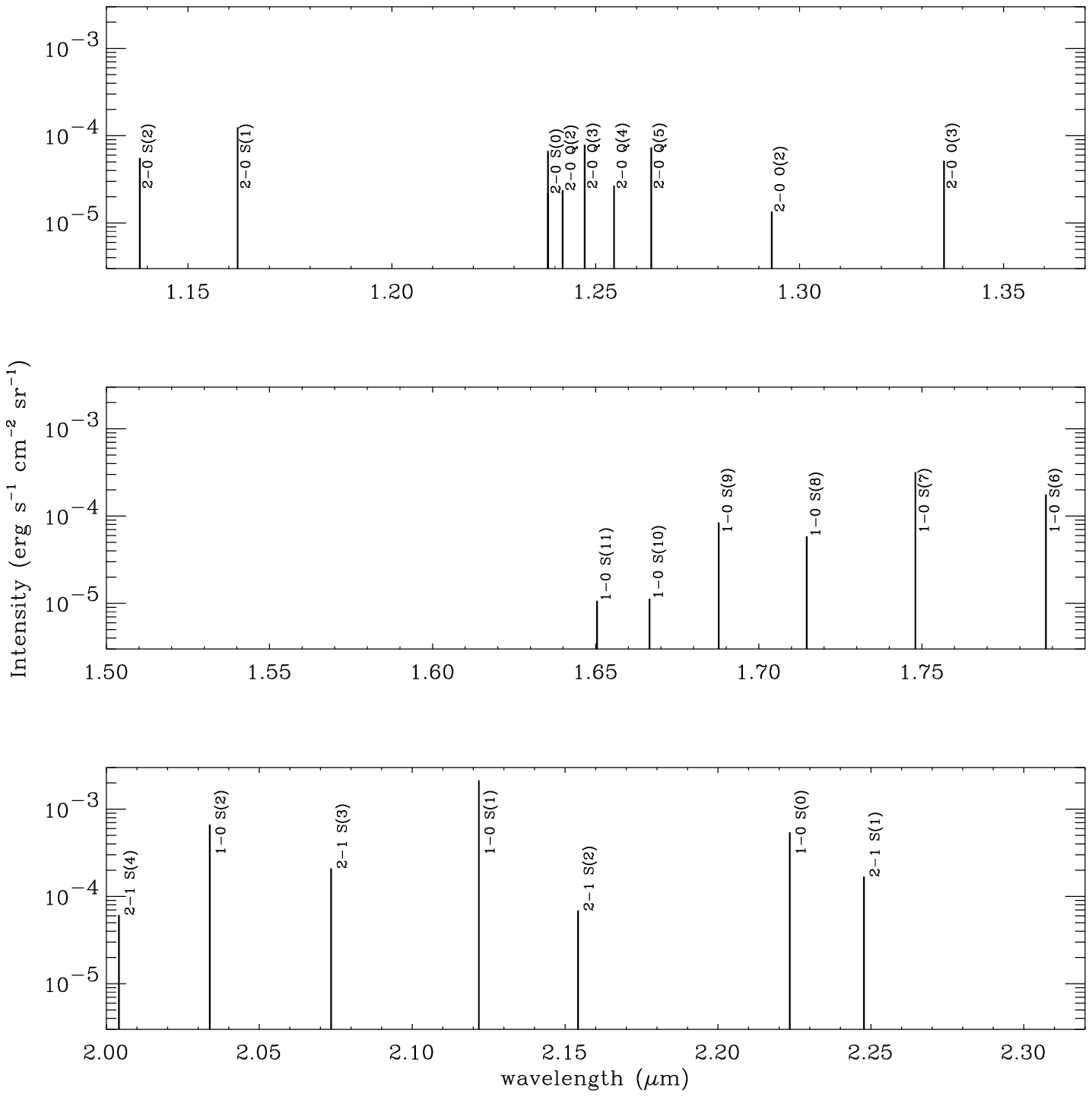,width=15cm}}

\vskip 1.5truecm
{Fig. 7b}
\end{figure}

\begin{figure}
\centerline{\psfig{figure=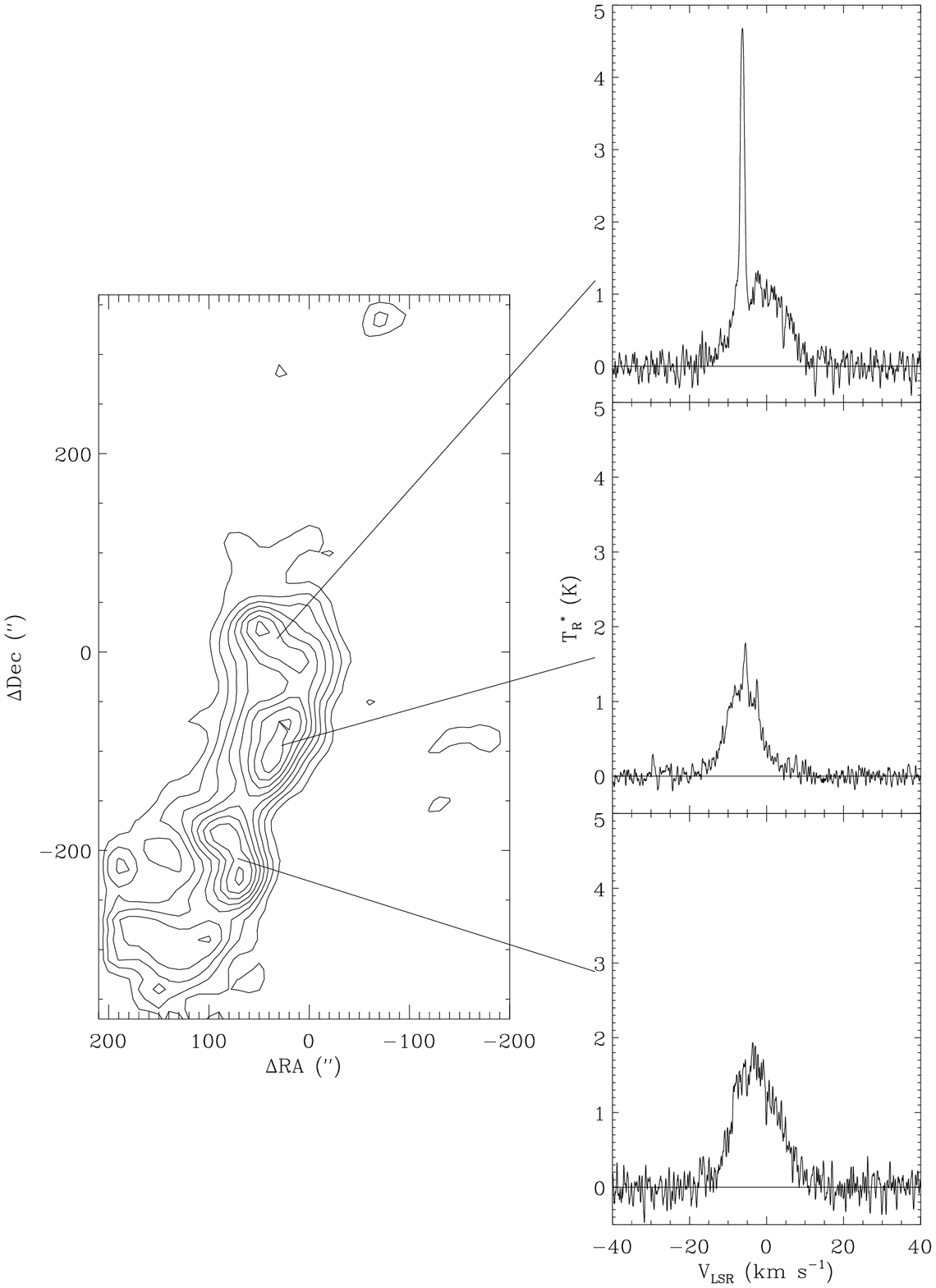,width=18cm}}

\vskip 0.5truecm
{Fig. 8}
\end{figure}


\clearpage
\begin{references}
\reference {} Arendt, R.G., 1989, ApJS, 70, 181

\reference{} Asaoka, I. \& Aschenbach, B. 1994, 284, 573

\reference {} Braun, R., \& Strom, R.G, 1986, Astr. Ap., 164, 193


\noindent Burton, M.G., Geballe, T.R, Brand, P.W., \& Webster, A.S., 1988, MNR
AS
, 231, 617

\reference {} Burton, Michael G., Hollenbach, D. J., Haas, M. R.,
Erickson, E. F., 1990, ApJ, 355, 197


\noindent Cesarsky, D, Cox, P., Pineau des Forets, G.P. et al., 1999, A\&A, 348, 945  

\noindent Cornett, R.H., Chin, G. and Knapp, G.R., 1977, ApJ, 54, 889

\reference{}  Cutri, R. M. et al., http://www.ipac.caltech.edu/2mass/releases/
first/doc/explsup.html

\reference{}Dwek, E., \& Arendt, R. G., 1992, ARA\&A, 30, 11

\reference{} Dickman, R. L., Snell, R. L.,
Ziurys, L. M., \& Huang, Y.-L. 1992, ApJ, 400, 203

\reference{} Draine, B. T., McKee, C.F., 1993, ARA\&A, 31, 373

\noindent Fesen, R.A, \& Kirshner, R.P., 1980, ApJ, 242, 1023

\reference {} Giovanelli, R., \& Haynes, M. P., 1979, ApJ, 230, 540


\noindent Graham, J.R., Wright, G.S., \& Longmore, ApJ, 1987, ApJ, 313, 847

\reference{} Hartigan, P.,Raymond, J., \& Hartmann, L., 1987, ApJ, 316, 323

\noindent Hollenbach, D. \& McKee, C. F., ApJ, 1989, 342, 306 (HM89)

\reference{} Innes, D.E., Giddings, J.R., Falle, S.A.E.G., 1987, MNRAS, 226, 6
71

\reference{} Mangum, J., 1997, http://www.tuc.nrao.edu/obsinfo.html

\reference {} McKee, C. F., 1987, $``$Spectroscopy of astrophysical 
plasmas", ed. by A. Dalgarno and D. Layzer.

\reference {} McKee, C. F., Chernoff, D. F., \& Hollenbach, D. J., 1984,
$``$Galactic and Extragalactic Infrared Spectroscopy", ed. by M. F. Kessler
 and J.P. Phillips

\reference {} Mufson, S. L. et al., 1986, AJ, 92, 1349

\reference {} Nugent et al., 1984, ApJ, 284, 612

\reference {} Nussbaumer, H., \& Storey, P.J., 1988, A\&A, 193, 327

\noindent Oliva,  E., Lutz, D., Drapatz, S., \& Moorwood, A. F.M., 1999a, A\&A,
 3
41, 75

\reference {} Oliva,  E., Moorwood, A. F. M., Drapatz, S., Lutz, D. \&
Sturm, E., 1999b, A\&A, 343, 943

\noindent Petre, R., Szymkiwiak, A.E., Seward, F.D., Willingale, R.,
1988, ApJ, 335, 215

\reference {} Reach, W. T., \& Rho, J., 2000, ApJ, in press (astro-ph/0007148)  

\reference {} Rho, J., Petre, R., and Hester, J.J., 1993, ``ROSAT
Observations of IC443'', at The Soft X-ray
Cosmos: ROSAT Science Symposium, ed. E.M. Schlegel, and R. Petre, p318

\noindent Richter, M.J, Graham, J.R., \& Wright, G. S., 1995, ApJ, 454, 277

\reference {} Rieke, G. H., \& Lebofsky, M.J., 1985, ApJ, 288, 618 

\reference {} Saken, J.M., Fesen, R.A. \& Shull, J.M. 1992, ApJS, 81, 715


\reference{} Skrutskie, et al., 1997, $``$The impact of large scale near-IR sky
surveys", ed.by F. Garzon et al., p25


\noindent van Dishoeck, E. F., Jansen, D.J, \& Phillips, T.G., 1993, A\&A, 279
,
541

\reference{} Wang, Z, \&  Scoville, N.Z., 1992, ApJ, 386, 158

\end{references}
\end{document}